\documentclass[%
 reprint,
amsmath,amssymb,
aps,
prb,
nofootinbib
]{revtex4-2}
\usepackage{amsmath}
\usepackage{dsfont}
\usepackage{physics}
\usepackage{amssymb}
\usepackage{graphicx}
\usepackage{float}
\usepackage{tabularx}
\usepackage{xcolor}
\usepackage{graphicx}% Include figure files
\usepackage{dcolumn}% Align table columns on decimal point
\usepackage{bm}% bold math
\usepackage{hyperref}

\usepackage{ulem}

\newcommand{\NFL}{\mathrm{NFL}}
\begin{document}

\preprint{APS/123-QED}

%\title{Rate Functions\\and the Approach to Adiabaticity in Quantum Many-Body Systems}% Force line breaks with \\
%\thanks{A footnote to the article title}%
\title{Quantum critical superconductivity in a dense fermi dilute bose mixture close to mechanical instability}
\author{Vibhu Mishra}
\email{vibhu.mishra@uni-goettingen.de}
% \altaffiliation[Also at ]{Physics Department, XYZ University.}%Lines break automatically or can be forced with \\

% \email{Second.Author@institution.edu}
\affiliation{%
 Institute for Theoretical Physics, Georg-August-Universität Göttingen, Friedrich-Hund-Platz 1, 37077 Göttingen,Germany
}%

%\collaboration{MUSO Collaboration}%\noaffiliation

% \homepage{http://www.Second.institution.edu/~Charlie.Author}
%\affiliation{Second institution and/or address\\ This line break forced% with \\}%
%\affiliation{ Third institution, the second for Charlie Author}%
%\author{Delta Author}
%\affiliation{% Authors' institution and/or address\\This line break forced with \textbackslash\textbackslash}%

%\collaboration{CLEO Collaboration}%\noaffiliation

\date{\today}% It is always \today, today,
             %  but any date may be explicitly specified

\begin{abstract}

We couple a metal to an uncondensed bose gas with repulsive bose-fermi interactions in two dimensions. 
Within the random phase approximation, we show that the density fluctuations of the bosons, which mediate the effective fermion-fermion attraction, become long ranged on appropriate tuning of interactions.
This leads to parametrically enhanced superconducting temperatures $T_c \propto g$ as opposed to $T_c \propto \exp(-1/g)$ for the BCS case.
The dilute bose gas density acts as the primary bottleneck to the enhancement.
We discuss the conceptual difficulties of implementing this idea in electron-exciton mixtures.

% \begin{description}
% \item[Usage]
% Secondary publications and information retrieval purposes.
% \item[Structure]
% You may use the \texttt{description} environment to structure your abstract;
% use the optional argument of the \verb+\item+ command to give the category of each item. 
% \end{description}
\end{abstract}

%\keywords{Suggested keywords}%Use showkeys class option if keyword
                              %display desired
\maketitle

%\tableofcontents

\section{Introduction}
\label{sec:intro}
%=========================================================================
The problem of investigating and enhancing superconductivity and in engineered quantum systems has gained significant interest in recent years \cite{Hofmann2022Heuristic, Esterlis_Bounds}.

A major line of attack on this problem involves engineering a scenario where, in the model Hamiltonian of the system, a fermion bilinear gets coupled to a boson bilinear as (all position-momentum, frequency-time, indices suppressed)
\begin{equation}
    H_I \sim \lambda \int \; c^\dagger c \; X^2,
\end{equation}
where $X \sim b + b^\dagger$ corresponds to some bosonic displacement field.
If we schematically write an effective fermion-fermion interaction with this starting point as
\begin{equation}
    H_{4c} \sim \lambda^2 \int \langle X^2 \cdot X^2 \rangle \; c^\dagger c^\dagger c c,
\end{equation}
we get a net attractive coupling strength that depends explicitly on the boson density which can then be controlled and modified to get an enhancement in $T_c$ \cite{Transient_Pumped}.
In a cleverer variant of the above scheme \cite{Resonantly_Enhanced}, a driving field is linearly coupled to some fermion hybridization between different levels $c$ and $d$ as $H_I \sim \lambda \int X \, c^\dagger d + \text{h.c}$.
$T_c$ can then be enhanced using resonance between the boson frequency and the energy gap between $c$ and $d$ levels.
For bosonic systems that violate particle number, the non-interacting bosonic ground state is just the empty vacuum and non-zero densities usually come from quantum or thermal fluctuations which are rather small.
For such cases it is necessary to drive the bosons externally and only then do we see appreciable enhancements in $T_c$.

A slightly adjacent line of attack involves coupling the fermions to bosons at finite density with interactions that conserve particle number as \cite{Exciton_Superconductor, Thermal_Boson_Paper}
\begin{equation}
    H_I \sim \lambda \int \; c^\dagger c \; b^\dagger b. 
\end{equation}
The obvious advantage is that a simple thermal bose distribution suffices even for a dilute bose gas to make significant enhancements to $T_c$.
This can be seen easily for weak couplings we have where we have
\begin{equation}
    T_c \propto e^{-1/g},
\end{equation}
with $0 < g \ll 1$ quantifying  the dimensionless fermion attraction.
A simple estimate\footnote{This is the reverse of the Gorkov-Melik-Barkhudarov (GMB) correction from induced interaction which is repulsive and for dilute Fermi gasses leads to $-g + g^2$ which essentially halves the original $T_c$ \cite{Pethick_Smith_Book, Pethick_GMB}.} shows that even a modest enhancement in attraction $-g \rightarrow -(g + g^2)$ gives  a terrific relative enhancement in the transition temperature as $e^{-(g + g^2)^{-1}} \approx e \; e^{-1/g} \implies  T_c \rightarrow e T_c$.
A nice analysis of such enhancements by coupling an $s$-wave superconductor to a bose gas was carried out in \cite{Thermal_Boson_Paper}.

If we now again schematically re-write the effective fermion interaction for such bose-fermi mixtures
\begin{equation}
    H_{4c} \sim \lambda^2 \int \langle b^\dagger  b \, b^\dagger b  \rangle \; c^\dagger c^\dagger \,c \, c,
\end{equation}
we immediately see that the fermions pair up with each other via the bose gas density fluctuations \cite{Density_Fluctuations}.
If we can tune the system in such a way that these fluctuations become long ranged then quantum critical pairing kicks in.
The crucial advantage here is that for quantum critical superconductivity, there is an algebraic dependence of $T_c$ on the pairing strength $g$ between the fermions as $T_c \propto g$ where $g$ needs to be figured out in terms of system parameters.
In three dimensions, the Hubbard-Stratonovic theory of Stoner magnetism shows a ferromagnetic instability when
$U \, \nu_F \sim 1,$ along with a metallic critical point \cite{altland_simons_2023}.
Here we look at something similar but for a bose-fermi mixture in two dimensions as enhancements to $T_c$ are stronger for lower dimensions.

%This problem is interesting as the investigations into the so called $\gamma$ model have shown that for quantum critical metals with long range attraction between fermions, the $T_c$ depends as a power law in the fermion-critical boson coupling strength and here we investigate the extent to which this temperature can be enhanced using the thermal boson gas.

The article is structured as follows, in Sec.[\ref{sec:Backdrop}] we provide further details that motivate this work.
Then in Sec.[\ref{sec:Model_System}], we discuss the specific model and write down the effective inter-particle interactions.
Then in Sec.[\ref{sec:Long_Ranged_Attraction}] we discuss the tuning condition to get long ranged fermion-fermion attraction.
In sections [\ref{sec:Inducing}] and [\ref{sec:Enhancing}] we look at $T_c$ when the bosons induce and enhance the pairing respectively.
In Sec.[\ref{sec:Bottlenecks}] we discuss the major physical roadblock to our theoretical study in the context of electron-exciton system. 
Finally we end in Sec.[\ref{sec:summary}] 

%=========================================================================
\section{Backdrop}
\label{sec:Backdrop}
%========================================================================
The textbook level study of Cooper instability generally always starts with the mean field gap equation analysis of the following BCS Hamiltonian
\begin{equation}
\label{eq:H_BCS}
    H_{\rm{BCS}} = \sum_{\mathbf{k} \sigma} \varepsilon_{\mathbf{k} \sigma} c_{\mathbf{k} \sigma}^{\dagger} c_{\mathbf{k} \sigma}+\sum_{\mathbf{k}, \mathbf{k}^{\prime}} V_{\mathbf{k}, \mathbf{k}^{\prime}} c_{\mathbf{k} \uparrow}^{\dagger} c_{-\mathbf{k} \downarrow}^{\dagger} c_{-\mathbf{k}^{\prime} \downarrow} c_{\mathbf{k}^{\prime} \uparrow},
\end{equation}
where in principle, the interaction matrix elements $V_{\mathbf{k}, \mathbf{k}^{\prime}}$ can be placed by hand to obtain results which can then be compared with experiments.
Then there is the separate task of coming up with an appropriate mechanism that generates the required $V_{\mathbf{k}, \mathbf{k}^{\prime}}$ of physical interest.

For isotropic homogeneous metals, we can label the interaction simply with the magnitude of momentum transfer $q \equiv |\mathbf{k} -\mathbf{k}^{\prime}|$ as $V_{\mathbf{k}, \mathbf{k}^{\prime}} = V_{|\mathbf{k} -\mathbf{k}^{\prime}|} = V_q$.
For systems with short range interactions we can now freely work with $V_{q=0} \equiv V$ and for small dimensionless interaction strength $g$ get the famous $T_c \sim \exp(-1/g)$ \cite{coleman2015introduction, bruus_Flensberg}.

This framework explains how an attractive interaction, however weak gives rise to Cooper instability below $T_c$.
Interestingly, it also shows quite robustly that for the case of attraction coming from electron-phonon coupling,  the retardation effects can be easily incorporated with a careful choice of $V_{\mathbf{k}, \mathbf{k}^{\prime}}$ which then shows that even  large Coulomb repulsion does not circumvent the onset of superconductivity in most cases \cite{coleman2015introduction, Sigrist2005Introduction}.

This standard framework breaks down for systems with strong electron phonon coupling and the whole problem has to be looked at afresh, which takes us to the Migdal-Eliashberg theory where the key technical change comes from the fact that the electron and phonon self energies have to be accounted for self consistently and in strong coupling limit it shows $T_c \propto \sqrt{g}$ for Einstein phonons \cite{mahan2000many, Marsiglio_Eliashberg, Eliashberg_Limits}.

An interesting but unrelated side remark has to be made.
If we look at the binding energy of a single Cooper pair evaluated in the simple Cooper problem \cite{coleman2015introduction, Leggett2006quantum}
\begin{equation}
    E_{\rm CP} 
    =
    \frac{\Lambda}{e^{1/g} - 1},
\end{equation}
we see that for $g \ll 1 \implies E_{\rm CP} \sim T_c \propto e^{-1/g}$ so they track nicely with each other whereas for $g \gg 1 \implies E_{\rm CP} \sim \Lambda g$ so there is a parametric mismatch between the $E_{\rm CP}$ and $T_c$.   

Even more interestingly, the Migdal-Eliashberg framework is still useful for long ranged dynamical interactions $V_{q, \Omega}$.
Such effective interactions routinely show up in the vicinity of metallic quantum critical points which almost always come together with higher than usual $T_c$.
Within this framework, the problem of solving the gap equations has been worked out in this century over the course of about 15 years by a thorough investigation of the so called $\gamma$ model \cite{GammaI, GammaII, GammaIII, GammaIV, GammaV, GammaVI} where $\gamma$ becomes the system specific object which distinguishes one class of critical metals from another.
%For this class of problems one has to simultaneously solve for the pairing gap $\Delta(i\omega)$ and the non-Fermi liquid fermion self energy $\Sigma(i\omega)$ which becomes the extra new ingredient to worry about compared to standard BCS workflow.
%The result relevant for our work is that $T_c \sim \mathfrak{g}$ where $\mathfrak{g}$ has the dimensions of temperature and depends polynomially on model parameters controlled by $z$.
There are important technicalities concerned with the pairing temperature $T_p$ and the proper superconducting temperature $T_{\rm{SC}}  \lesssim  T_p$ but we do not stress over these distinctions here \cite{Chubukov2020_interplay} and work with a single $T_c$.

An important question which is not directly addressed there is the influence of the original Coulomb repulsion which is assumed to be small.
This is important because for such systems, there's no Debey frequency which is much smaller than the fermion bandwidth so there's no analogue of a retardation effect to the rescue.

The reason an attraction peaked\footnote{For a simple case $V_q \sim [q^2 + \xi^{-2}]^{-1}$ is peaked at $q=0$ as opposed to $V_q \sim [(q - Q)^2 + \xi^{-2}]^{-1}$ which peaks at some ordering vector $q = Q$.} at $q=0$ is particularly useful
for superconductivity can be seen directly from the patch decomposition renormalization group (RG) analysis of
Ref.~\cite{Metlitski_Cooper_Pairs}.  
For an isotropic Fermi surface, the BCS interaction can be decomposed into angular momentum channels,
\begin{equation}
V(\theta-\theta')
=
\sum_j V_j e^{ij(\theta-\theta')}.
\end{equation}
The exchange of a critical boson with momentum $q\ll k_F$ generates an
additional attractive contribution to the flow of $V_j$.  Since the angular
factor associated with this small momentum transfer satisfies
$\cos(jq/k_F)\simeq1$, this contribution $\mathcal A>0$ is almost independent of $j$, and schematically the flow takes the form
\begin{equation}
\frac{d\widetilde V_j}{d\ell}
=
-\widetilde V_j^2-\mathcal A .
\end{equation}
Thus the long ranged fluctuations drive every pairing channel towards stronger attraction, including channels in which $\widetilde V_j$ is repulsive.  
The different channels still begin with different $\widetilde V_j$ and consequently the one which flows to strong coupling first, sets the pairing channel and controls the quantitative value of $T_c$.
However, isotropy is not essential for the underlying forward scattering enhancement, although the exact equality between angular momentum channels no longer applies on a generic Fermi surface \cite{Berg_anisotropic_Nematic, Chubukov_anisotropic_Nematic}.

Both the $\gamma$ model \cite{Chubukov2020_interplay, Patel_Sachdev_Large_N, Esterlis_Critical_Eliashberg} and direct RG \cite{Metlitski_Cooper_Pairs} analysis show that for the case of quantum critical pairing, up to pre-factors we have the extremely useful result
\begin{equation}
    T_c \sim T_{\rm NFL}.
\end{equation}
For a quantum critical metal $T_{\NFL}$ is defined as the energy scale where the fermion self energy $\Sigma(i\omega)$ becomes comparable to its bare kinetic frequency term $i\omega$ so schematically $\Sigma(T_{\rm NFL}) \simeq T_{\rm NFL}$.
Now for a given $V_{q, \Omega}$, it's extremely easy to evaluate $\Sigma(i\omega)$ and consequently $T_{\rm NFL}$ using standard methods \cite{Sachdev_Phases, Sachdev_QPT}.

In this work we aim to exploit these two facets of quantum criticality with the hopes of enhancing $T_c$ namely (i) the algebraic rather than exponential dependence of $T_c$ on coupling strength and (ii) the enhancement of superconductivity in any and all pairing channels due to $V_{q, \Omega}$ that peaks at $q=0$.
We engineer such a scenario in a bose-fermi mixture to obtain the appropriate $V_{q, \Omega}$ which then directly provides us with $\Sigma(i\omega) $ which immediately gives $ T_{\rm NFL} \sim T_c$.
In this sense the current work is simply an elaborate matching between the model parameters and corresponding $T_{\rm NFL}$ which we believe to provide a nice zeroth-order insight into the enhancement of $T_c$ for such systems.

%We try to engineer such a $V_{q, \Omega}$ as an effective interaction between electrons by coupling a two dimensional metal with an excitonic gas which we investigate in the rest of the work.

A brief outline of the strategy is as follows.
As discussed up above, the BCS analysis of $H_{\text{int}} \sim -g \sum c^{\dagger}c^{\dagger} cc$, where the attraction lives within an energy window $\Lambda$ around the Fermi surface immediately gives $T_c \propto \Lambda \exp(-1/g)$.
Starting from a more microscopic electron-phonon system (Fröhlich/Holstein model etc) then simply involves identifying the energy window as the Debey frequency $\Lambda \sim \omega_D$ and relating $g$ to the appropriate electron-phonon coupling and Fermi surface density of states $\nu_F$.
Similarly here, we start with a bose-fermi mixture and write down an effective
$H_{\text{int}} \sim \sum V_{q, \Omega} c^{\dagger}c^{\dagger} cc$ and look for the condition when $V_{q, \Omega}$ becomes long ranged which turns out to be that of mechanical instability.

If the BCS metal is coupled to an additional bath that gives rise to an effective $H_{\rm int} \sim -g' \sum c^{\dagger}c^{\dagger} cc$ with the same interaction bandwidth $\Lambda$ then the couplings simply add up and the enhancement can be seen directly from
\begin{equation}
    T_c \sim \Lambda \exp\left(-\frac1{g+g'}\right)
\end{equation}
For the realistic case of different $\Lambda$'s, the enhancement still takes place however it can't be expressed in such a simple formula and the problem has to be worked out from appropriate $V_{\textbf{k}, \textbf{k}'}$ in Eq.\eqref{eq:H_BCS}.

In exactly the same vein, if the induced long ranged attraction from the bose gas $V_{\rm BG}$ is added on top of a quantum critical superconductor with its own long range $V_{\rm C}$, this also leads to enhancement which in generic cases has to be worked out from a full solution of the $\gamma$ model gap equations with combined $V_{\rm TOT} = V_{\rm BG} + V_{\rm C}$ which lies much outside the scope of this work.
The most trivial way to see a direct enhancement is if $\gamma$ for both $V_{\rm BG}$  and $V_{\rm C}$ are the same as that again allows for a direct addition of the attraction strengths in $V_{\rm TOT}$ to give the enhanced 
\begin{equation}
\label{eq:Enhancement_Concrete}
    T_c \sim \left[\mathfrak{g}^{\gamma} + (\mathfrak{g}')^{\gamma} \right]^{1/\gamma},
\end{equation}
where $\mathfrak{g}$ and $\mathfrak{g}'$ are the $T_{\rm NFL}$ are from  $V_{\rm C}$ and $V_{\rm BG}$ respectively and the above formula tells us how the temperatures combine together when the $\gamma$'s are the same.
The non-trivial part here doesn't come from demonstrating the enhancement which is self evident in this artificial case but in a careful evaluation of $z, \gamma$ and $\mathfrak{g}'$ for $V_{\rm BG}$.
The crucial difference comes from the fact that $V_{\rm BG}$ is controlled by the inverse dynamical bose susceptibility which shows strong $T$ dependence and gives $\Omega^2/q^2$ as opposed to $\Omega/q$ in conventional critical metals.

\subsection{$d, z$ and $\gamma$}
\label{subsec:Exponents}

We quickly run through the relations between the various quantities characterizing a critical metal \cite{Sachdev_Phases, Sachdev_QPT, Chubukov2020_interplay}.
Critical interaction $V_{q, \Omega} \propto \left[\frac{\Omega^\alpha}{q^\beta}+q^\delta \right]^{-1}$ immediately gives the dynamical critical exponent $\Omega \sim q^z$ as
\begin{equation}
    z = \frac{\beta + \delta}{\alpha}.
\end{equation}

To obtain $\gamma$ in $d$ spatial dimension, we use the definition $V_\Omega \propto \int q^{d-2} V_{q, \Omega} \, d q \propto \Omega^{-\gamma}$ \cite{Chubukov2020_interplay} which on evaluating
\begin{equation}
    \int \frac{q^{d-2} d q}{\frac{\Omega^\alpha}{q^\beta}+q^\delta}
    =
    \frac{\Omega^{\frac{d + \beta - 1}{\beta + \delta} \alpha}}{\Omega^\alpha}
    \int d t \frac{ \, t^{\frac{d + \beta - 1}{\beta + \delta}}}{t+1} 
    \sim
    \Omega^{-\gamma}
\end{equation}
gives
\begin{equation}
    \gamma 
    = 
    \frac{\alpha(1 + \delta - d)}{\beta + \delta}  
    = 
    \frac{1 + \delta - d}{z},
\end{equation}
which shows that $\gamma$ is higher in lower $d$.
Finally the fermion self energy in relation to $\gamma$ is
\begin{equation}
    \Sigma(i\omega) 
    \propto  
    \int_0^{|\omega|} \Omega^{-\gamma} \, d\Omega
    \sim \mathfrak{g}^{\gamma}|\omega|^{1-\gamma}.
\end{equation}
Using $T_{\rm NFL} \sim \Sigma(T_{\rm NFL})$ here we can immediately infer
\begin{equation}
    T_{\rm NFL} \sim \mathfrak{g}.
\end{equation}

We show later that for $V_{\rm BG}$, $d = \alpha = \beta = \delta = 2$ where these exponents come purely from the inverse susceptibility of a free dilute bose gas and give $z = 2$ and $\gamma = 1/2$ so the same exponents as a CDW/SDW metal but with zero ordering vector.

The case of $(d,\delta) = (2, 1)$ immediately gives $\gamma = 0$ and corresponds to a marginal-Fermi liquid \cite{MFL} which, as the name suggests is almost a Fermi liquid and the parametric enhancements we center this work around are basically washed away.
A closer look at the $q$ dependence of the electron-exciton and exciton-exciton interaction makes $\delta=1$ a reality at small $q$ and we discuss this more in Sec.[\ref{sec:Bottlenecks}].

\section{Model System}
\label{sec:Model_System}

\begin{figure}
	\centering
	 %\begin{subfigure}
		\includegraphics[scale=1.0]{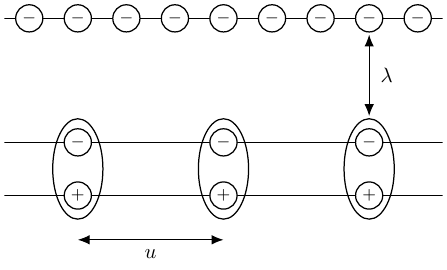}
		% \caption{Yellow color}
		%\label{fig:subfigA}
	 %\end{subfigure}
	\caption{The model system of dense electrons in top layer interacting with dilute excitons below that motivate the field theory under study.
    Both $\lambda$ and $u$ are repulsive due to the orientation of the excitonic dipole.
    The fermion boson interaction is repulsive as opposed to the cases looked at in \cite{Thermal_Boson_Paper, Exciton_Superconductor} which circumvents electron-exciton bound states.
    The drawback of this specific physical setup for our purposes is discussed in Sec.[\ref{sec:Bottlenecks}].}
	\label{fig:ATP}
\end{figure}

We investigate the following action \cite{Thermal_Boson_Paper, Exciton_Superconductor} motivated from the physical system in Fig.[\ref{fig:ATP}]
\begin{equation}
    S = S_c  + S_b  + S_{bb} + S_{cb},
\end{equation}
with ($\bar{k}\equiv(\mathbf k,i\omega_n),\quad
\bar{p}\equiv(\mathbf p,i\nu_m),\quad
\bar{q}\equiv(\mathbf q,i\Omega_m)$)
\begin{equation}
    \begin{aligned}
        S_c &= 
         \int_{\bar{k}} c_{\bar{k}}^{\dagger} \left[ -i\omega_n + \varepsilon_{\textbf{k}} \right] c_{\bar{k}} 
        \\
        S_b &= 
         \int_{\bar{p}} b_{\bar{p}}^{\dagger} \left[ -i\nu_m + \xi_{\textbf{p}} \right] b_{\bar{p}} 
        \\
        S_{cb} &= 
        \lambda \int_{\bar{k}} \int_{\bar{p} } \int_{\bar{q}}  c_{\bar{k} + \bar{q}}^{\dagger} \, c_{\bar{k}} \; b_{\bar{p} - \bar{q}}^{\dagger} \, b_{\bar{p}} 
        \\
        S_{bb} &= 
        \frac{u}{2} \int_{\bar{k}} \int_{\bar{p} } \int_{\bar{q}}  b_{\bar{k} + \bar{q}}^{\dagger} \, b_{\bar{k}} \; b_{\bar{p} - \bar{q}}^{\dagger} \, b_{\bar{p}}
        \; 
    \end{aligned}
\end{equation}
where both $u$ and $\lambda$ are both repulsive so positive.

%Before proceeding with the analysis of this QFT we have to point out that for the actual physical exciton repulsion, $u \to u(q) = u_0 + O(q)$ so in this work we are basically working with the direct contact piece $u_0$.
%As we will show in Sec.[\ref{sec:Bottlenecks}], the $O(q)$ piece, which we by and large neglect in this work is actually genuinely problematic and can only be patched up by working with $\lambda \to \lambda(q)$ which complicates the actual physics.
%In the bulk of this work we'll ignore these complexities to obtain crude but illuminating insights before going back to the issues at hand.

We work with parabolic dispersion for bosons $\xi_{\textbf{p}} = p^2/2M  - \mu$ with density $n_b$ and characterize the fermions only with quantities defined locally on the Fermi surface (so the Fermi momenta $k_F$ is absent) like the Fermi velocity $v_F$, surface curvature $\kappa_F$, density of states $\nu_F$ and its derivatives $\nu_F', \nu_F''$ all of which in principle can be evaluated for a given $\varepsilon_{\textbf{k}}$ as given in Appendix.[\ref{app:fermion_lindhard_general}] and are angle independent for an isotropic system.
The reason for working this way is that the free fermion bubble plays crucial role for the critical properties while the free Fermi gas for $d=2$ has a constant $\nu_F$ as a consequence the small $q$ static Lindhard bubble is independent of $q$ \cite{Mihaila_Lindhard} which is too strong an approximation for realistic systems. 

The shorthand integration means
\begin{equation}
    \int_{\bar{k}} \equiv T\sum_{\omega_n}
\int\frac{d^2 k}{(2\pi)^2}.
\end{equation}

We start by expressing the interaction as a matrix product
\begin{equation}
    S_{i n t}=\frac{1}{2} \int_{\bar{q}} \boldsymbol{\rho}^T(\bar{q}) \, \textbf{V}_0 \, \boldsymbol{\rho}(-\bar{q}), \quad \textbf{V}_0 = \left(\begin{array}{ll}
0 & \lambda \\
\lambda & u
\end{array}\right)
\end{equation}
which has been expressed in terms of
\begin{equation}
    \rho_f(\bar{q}) \equiv \int_{\bar{k}} c_{\bar{k}+\bar{q}}^{\dagger} c_{\bar{k}}, \quad \rho_b(\bar{q}) \equiv \int_{\bar{p}} b_{\bar{p}+\bar{q}}^{\dagger} b_{\bar{p}},
\end{equation}
 with row vector $\boldsymbol{\rho}^T(\bar{q})  \equiv \left(\rho_f(\bar{q}), \rho_b(\bar{q})\right)$.
%For this work we explicitly need any dispersion other than the usual parabolic $\varepsilon_{\textbf{k}} = k^2/2m - \mu_F$ as for this case the static Lindhard bubble is independent of the momentum transfer $q$ and this is undesirable (this problem is not there in  dimensions though so there a spherical dispersion works).

%We work with simple spherical dispersions for both fermions $\varepsilon_{\textbf{k}} \equiv \frac{k^2}{2m} - \mu_c$ and bosons $\xi_{\textbf{p}} = \frac{p^2}{2M} - \mu$ with the notation $k \equiv |\textbf{k}|$.

We need all our relevant physical quantities in terms of the free fermion and free boson polarization bubble
\begin{equation}
    \begin{aligned}
\Pi_f\left(\bar{q}\right) & \equiv \int \frac{d^2 k}{(2 \pi)^2} \frac{n_F\left(\varepsilon_{\textbf{k}+\textbf{q}}\right)-n_F\left(\varepsilon_\textbf{k}\right)}{i \Omega_m-\varepsilon_{\textbf{k}+\textbf{q}}+\varepsilon_\textbf{k}} 
\\
\Pi_b\left(\bar{q}\right) & \equiv \int \frac{d^2 p}{(2 \pi)^2} \frac{n_B\left(\xi_{\textbf{p}+\textbf{q}}\right)-n_B\left(\xi_\textbf{p}\right)}{i \Omega_m-\xi_{\textbf{p}+\textbf{q}}+\xi_\textbf{p}}. 
\end{aligned}
\end{equation}
Within the random phase approximation (RPA) we write down the dressed inverse polarization matrix as
\begin{equation}
    \boldsymbol{\Pi}^{-1}(\bar{q})
    = 
    \boldsymbol{\Pi}_0^{-1}(\bar{q}) + \textbf{V}_0 = \left(\begin{array}{cc}
\Pi_f^{-1}(\bar{q}) & \lambda \\
\lambda & \Pi_b^{-1}(\bar{q})+u
\end{array}\right).
\end{equation}
where $\boldsymbol{\Pi_0}(\bar{q}) \equiv \left(\begin{array}{cc}
\Pi_f(\bar{q}) & 0 \\
0 & \Pi_b(\bar{q})
\end{array}\right)$
\cite{Coulomb_Drag, Coulomb_Drag_Graphene, Near_Field_Coulomb_Drag}.

Inverting the dressed polarization matrix gives us \cite{Collisionless_Sound, Collective_Modes, Phonon_Exchange, 2D_Bose_Gas_Normal_State}
\begin{equation}
    \begin{aligned}
        \Pi_{f f}(\bar{q}) &=\frac{\Pi_f(\bar{q})}{1-\lambda^2 \Pi_f(\bar{q}) \tilde{\Pi}_b(\bar{q})}, 
        \\ 
        \Pi_{b b}(\bar{q}) &=\frac{\tilde{\Pi}_b(\bar{q})}{1 - \lambda^2 \Pi_f \tilde{\Pi}_b(\bar{q})}, 
        \\ 
        \Pi_{f b}(\bar{q}) &= -\frac{\lambda \Pi_f(\bar{q}) \tilde{\Pi}_b(\bar{q})}{1 - \lambda^2 \Pi_f(\bar{q}) \tilde{\Pi}_b(\bar{q})} 
        =
         \Pi_{b f}(\bar{q}),
        \\
        \tilde{\Pi}_b(\bar{q}) &\equiv \frac{\Pi_b(\bar{q})}{1+u \Pi_b(\bar{q})}.
    \end{aligned}
\end{equation}

The dielectric matrix $\boldsymbol{\varepsilon}(\bar{q})
        \equiv \mathds{1} + \textbf{V}_0 \boldsymbol{\Pi_0}(\bar{q}),$ gives us the dressed effective interaction as
\begin{equation}
    \textbf{V}(\bar{q})
    \equiv
        \boldsymbol{\varepsilon}^{-1}(\bar{q}) \textbf{V}_0
        =
    \frac{1}{\text{D}(\bar{q})}\left(\begin{array}{cc}
-\lambda^2 \Pi_b(\bar{q}) & \lambda \\
\lambda & u-\lambda^2 \Pi_f(\bar{q})
\end{array}\right).
\end{equation}
where \cite{RPA_V_Source}
\begin{equation}
    \begin{aligned}
        %=\textbf{V}-\textbf{V} \boldsymbol{\chi}(\bar{q}) \textbf{V}, 
        \\ \text{D}(\bar{q}) &\equiv \operatorname{det} [\boldsymbol {\varepsilon}(\bar{q})]
        =1+u \Pi_b(\bar{q})-\lambda^2 \Pi_f(\bar{q}) \Pi_b(\bar{q})
        \\
        &=\left(1+u \Pi_b(\bar{q})\right)\left(1-\lambda^2 \Pi_f(\bar{q}) \tilde{\Pi}_b(\bar{q})\right).
    \end{aligned}
\end{equation}

Now we can finally write down the individual effective interaction pieces
\begin{equation}
    \begin{aligned}
        V_{f f}(\bar{q})&= -\frac{\lambda^2 \tilde{\Pi}_b(\bar{q})}{1 - \lambda^2 \Pi_f(\bar{q}) \tilde{\Pi}_b(\bar{q})} 
        = -\lambda^2 \Pi_{b b}(\bar{q}), 
        \\ 
        V_{f b}(\bar{q})&=\frac{\lambda}{1+u \Pi_b(\bar{q})-\lambda^2 \Pi_f(\bar{q}) \Pi_b(\bar{q})}, 
        \\ 
        V_{b b}(\bar{q})&=\frac{u-\lambda^2 \Pi_f(\bar{q})}{1+u \Pi_b(\bar{q})-\lambda^2 \Pi_f(\bar{q}) \Pi_b(\bar{q})}.
    \end{aligned}
\end{equation}

The $V_{ff}(\bar{q})$ we see is almost the effective fermion-fermion interaction from \cite{Pethick_Smith_Book, Pethick_GMB} but with an extra $1 - \lambda^2 \Pi_f(\bar{q}) \tilde{\Pi}_b (\bar{q})$ in the denominator which makes all the difference in what follows.

\section{Long range attraction}

\label{sec:Long_Ranged_Attraction}

Zooming in on the dressed fermion-fermion interaction
\begin{equation}
    V_{f f}(\bar{q}) =\frac{-\lambda^2 }{\Pi_b^{-1}(\bar{q}) + u -\lambda^2 \Pi_f(\bar{q}) },
\end{equation}
we see the static piece in the denominator gives us the effective range of the interaction
\begin{equation}
    \Xi(T) = \Pi_b^{-1}(0) + u -\lambda^2 \Pi_f(0)
\end{equation}
where the dominant $T$ dependence comes from $\Pi_b^{-1}(0)$.
%Technically $\Pi_f$ also depends on $T$ but there the corrections over $T=0$ go as $T/T_F$ which are vanishingly small for our system.

For usual quantum critical metals criticality is defined by the following condition \cite{Sachdev_Phases, Sachdev_QPT}
\begin{equation}
    \Xi(T=0) = 0.
\end{equation}

As shown in Appendix.[\ref{app:Bose_Lindhard}] for the free bose gas we have
\begin{equation}
\Pi_b(0)
=
\frac{M}{2\pi}
\left[
\exp\left(
\frac{2\pi n_b}{MT}
\right)-1
\right].
\end{equation}

so at $T=0$ we have $\Pi_b^{-1}(0) = 0$ giving us the required tuning to obtain long range fermion-fermion  attraction as 
\begin{equation}
    u - \lambda^2 \Pi_f(0) = 0,
\end{equation}
which  immediately gives us
\begin{equation}
    \lambda^2 = u \ \nu_F^{-1},
\end{equation}
as the condition for criticality and from now on we substitute this value of $\lambda^2$ throughout.
Note that this is the saturation of the same stability bound shown in \cite{Mechanical_instability, Pethick_Smith_Book} if we plug in a $d=2$ Fermi gas dispersion there. 
A heuristic understanding can be obtained by picturing that the original boson-boson repulsion $u$ is offset by fermion induced effective boson-boson attraction $\lambda^2 \nu_F$.

At this point the extremely useful and fortunate fact is that for this parameter tuning, $V_{bb}$ and $V_{bf}$ actually do \textit{not} get long ranged as can be seen by the fact that for the above condition
\begin{equation}
    \text{D}(\bar{q} = 0) = 1,
\end{equation}
so the static piece never cancels off and on top of that
\begin{equation}
    V_{bb}(\bar{q} = 0) = 0.
\end{equation}

The fermion polarization bubble for $|\Omega_m| \ll v_F q$ is
\begin{equation}
\Pi_f(\bar{q})
=
\nu_F
+
C_Fq^2
-
\frac{\nu_F}{v_F}
\frac{|\Omega_m|}{q},
\end{equation}
where\footnote{$\kappa_F$ is the Fermi surface curvature not the Fermi momenta $k_F$.}
\begin{equation}
    C_F
    =
    \frac{\kappa_F\nu_F'}{12}
+
\frac{v_F^2}{24}
\left[
\nu_F''
-
2\frac{(\nu_F')^2}{\nu_F}
\right],
\end{equation}
the derivation of which can be found in Appendix.[\ref{app:fermion_lindhard_general}].
For a parabolic dispersion $\varepsilon_{\textbf{k}} = k^2/2m_f - \mu_f$ we have a constant density of states $\nu_F = m_f/2\pi  \implies C_F = 0$ and there is no $q^2$ dependent piece for small $q$ \cite{Mihaila_Lindhard}.

An important question we need to address before proceeding further is whether the bose gas self energy $\Sigma_b$ get drastically modified due to the critical Fermi surface.
In Appendix.[\ref{app:Boson_Self_Energy}] we show that within the conventional $T$-matrix approximation, $\Sigma_b$ largely doesn't care whether the Fermi surface is critical or not as within the patch approximation the dependence of $\Sigma_b$ on fermion self energy $\Sigma(i \omega)$ drops out entirely.

The functional form of the boson bubble $\Pi_b(\bar{q})$ changes dramatically depending on whether we are above or below the quantum degeneracy temperature $T_b$ defined as
\begin{equation}
    T_b \equiv \frac{2\pi n_b}{M}.
\end{equation}

%=========================================================================

\section{Inducing Superconductivity in Fermi Liquid}
\label{sec:Inducing}

If the original metal is a Fermi liquid then due to Hubbard/Coulomb repulsion, superconductivity is possible only via the Kohn-Luttinger mechanism which generically leads to absurdly low $T_c$ \cite{Kohn_Luttinger_Robustness, Fu_Kohn_Luttinger} and the bose gas essentially induces superconductivity at a temperature $T_{\rm IND}$.
In this section we assume that $T_b \gg T_{\rm IND}$ which immediately sets the boson dynamical bubble \cite{Collisionless_Sound}
\begin{equation}
\label{eq:Low_T:Boson_Bubble_Inverse}
\Pi_b^{-1}(\bar{q})
=
\frac{q^2}{4Mn_b}
+
\frac{M\Omega_m^2}{n_bq^2}.
\end{equation}
and now at criticality we can write down the effective fermion-fermion attraction as
\begin{equation}
\label{eq:Effective_ff_Low_T}
     V_{ff}(\bar{q}) 
    = \frac{-2u n_b/\nu_F}{2M\frac{\Omega_m^2}{q^2} + \frac{2u n_b}{v_F}\frac{ \left|\Omega_m\right|}{  q} + \frac{q^2}{2M} \left( 1 - \frac{4u n_b M C_F}{\nu_F}   \right)}
\end{equation}

To evaluate the self energy we work locally around a point $\mathbf k_F$ on
the Fermi surface and decompose $\textbf{q}$ into $q_\perp$ which is along and  $q_\parallel$ which is orthogonal to $\mathbf k_F$.  
The low energy patch scaling gives
$q_\perp\sim q_\parallel^2$, so that $q_\perp^2$ is subleading in
$V_{ff}$ and its momentum dependence can be evaluated at $q\simeq |q_\parallel|$ while $q_\perp$ remains important in the fermion propagator.
We start with the fermion self energy 
\begin{equation}
    \Sigma(i \omega)=-\int \frac{d \Omega}{2 \pi} \int \frac{d q_\perp d q_\parallel}{(2 \pi)^2} V_{f f}\left(q_\parallel, i \Omega\right) G\left(\mathbf{k}_F+\mathbf{q}, i \omega+i \Omega\right)
\end{equation}
which can be evaluated in a straightforward manner using the methods in \cite{Sachdev_QPT}.
Starting with the linearized dispersion round $\mathbf k_F$
\begin{equation}
    \varepsilon_{\mathbf{k}_F+\mathbf{q}} \simeq v_F q_\perp+\frac{\kappa_F }{2}q_\parallel^2,
\end{equation}
we first integrate over the transverse momenta $q_\perp$
\begin{equation}
\begin{aligned}
        \int d q_\perp G\left(\mathbf{k}_F+\mathbf{q}, i \omega+i \Omega\right)
    &= 
    \int  \frac{d q_\perp}{W(i\omega+i\Omega)-v_F q_\perp-\frac{\kappa_F }{2}q_\parallel^2}
    \\
    &=
    -\frac{i \pi}{v_F} \operatorname{sgn}(\omega+\Omega),
\end{aligned}
\end{equation}
where $W(i\omega) \equiv i\omega - \Sigma(i\omega)$  and we use
$\operatorname{sgn}\left[\omega-\frac{\Sigma(i \omega)}{i}\right]=\operatorname{sgn}(\omega)$ \cite{Sachdev_Phases}
to obtain
\begin{equation}
    \Sigma(i \omega)
    =
    -\frac{ i \Gamma}{2 \nu_F} \int_{-\infty}^{\infty}\int_{-\infty}^{\infty} \frac{d \Omega \, d q_\parallel}{(2 \pi)^2} \frac{\operatorname{sgn}(\omega+\Omega)}{c^2q_\parallel^2 + \Gamma \frac{  \left|\Omega\right|}{ q_\parallel} + 2M\frac{\Omega^2}{q_\parallel^2}}
\end{equation}
where 
\begin{equation}
    c^2\equiv \frac{1}{2M} \left( 1 - \frac{4u n_b M C_F}{\nu_F}   \right),
    \qquad
\Gamma\equiv\frac{2un_b}{v_F}.
\end{equation}

%Now we simply have to evaluate $T_{\rm NFL}$ using $\Sigma(T_{\rm{NFL}}) \sim T_{\rm{NFL}} .$

We immediately see though that $\Sigma(i\omega)$ is relatively non-trivial with both $z=3$ and $z=2$ pieces.
We use $\partial_{\omega}
\operatorname{sgn}(\omega+\Omega)
=
2\delta(\omega+\Omega),$
to collapse the frequency integral and obtain
\begin{equation}
\partial_{\omega}\Sigma(i\omega)
=
-i\frac{2\Gamma}{\pi^2\nu_F}\,
I(|\omega|),
\end{equation}
with
\begin{equation}
I(\omega)
\equiv
\int_0^\infty 
\frac{dq}
{c^2q^2+\Gamma\omega/q+2M\omega^2/q^2}.
\end{equation}

Anticipating the fact that at small enough $T$ the physics is still controlled by $z=3$ fermionic Landau damping we introduce the  dimensionless momentum $q = q_3(\omega)x$ in terms of which we write
\begin{equation}
I(\omega)
=
(c^2)^{-2/3}\Gamma^{-1/3}\omega^{-1/3}
F(\rho),
\end{equation}
where
\begin{equation}
\label{eq:F(r)_definition}
F(\rho)
\equiv
\int_0^\infty
\frac{dx}{x^2+x^{-1} + \rho x^{-2}},
\quad
q_3(\omega)
\equiv
\left(
\frac{\Gamma\omega}{c^2}
\right)^{1/3},
\end{equation}

and
\begin{equation}
\rho
\equiv
\frac{2M(c^2)^{1/3}}{\Gamma^{4/3}}\,
\omega^{2/3}
=
\left(
\frac{\omega}{\Theta}
\right)^{2/3}.
\end{equation}

which gives the crossover temperature scale
\begin{equation}
\Theta
=
\frac{\Gamma^2}
{(2M)^{3/2}c}
=
\frac{\sqrt{2}(un_b/v_F)^2}
{M^{3/2}c}.
\end{equation}
So for $T\gg\Theta \implies \rho\gg1$ (on going to the Matsubara frequencies $\omega_n$), where the $z=2$ Bose-gas dynamics dominates, whereas for $T \ll \Theta \implies \rho \ll 1$, it's $z=3$ fermionic Landau damping that dominates.

Since $\Sigma(0)=0$, integrating back in frequency gives
\begin{equation}
\Sigma(i\omega)
=
-i\,\operatorname{sgn}(\omega)\,
\Theta \, g
\, 
H\left(
\frac{|\omega|}{\Theta}
\right),
\end{equation}
where
\begin{equation}
H(y)
\equiv
\int_0^y
dt\,
t^{-1/3}
F(t^{2/3}),
\end{equation}
and the dimensionless crossover coupling is
\begin{equation}
g
\equiv
\frac{\sqrt{2M}}
{2\pi^2\nu_F c} 
\gg 1.
\end{equation}
for excitons much heavier than electrons.

The asymptotic behavior of $F(\rho)$ follows directly from the
dominant terms in its denominator,
\begin{equation}
F(\rho)
=
\int_0^\infty dx\,
\frac{x^2}{x^4+x+\rho}.
\end{equation}

For $\rho \ll 1$, we can drop $\rho$ altogether since $F(\rho)$ is nearly independent of $\rho$.

For $\rho \gg 1$, the dominant momentum scale is determined by setting $x^4 \sim \rho,$ (this can be checked by minimizing the denominator of Eq.\eqref{eq:F(r)_definition}) so we rescale $x = \rho^{1/4}y$
which gives
\begin{equation}
F(\rho)
=
\rho^{-1/4}
\int_0^\infty dy\,
\frac{y^2}
{y^4+ 1 + \rho^{-3/4}y}.
\end{equation}
and we ignore the last term in the denominator to get
\begin{equation}
F(\rho \gg 1)
\propto
\rho^{-1/4}
\end{equation}

%Thus $F(\kappa)$ approaches a constant in the Landau-damped$\kappa\ll1$ regime, while it scales as $\kappa^{-1/4}$ when the intrinsic bosonic dynamics dominates at $\kappa\gg1$.

On placing back the asymptotic behavior of $F(\rho)$ we can see that
\begin{equation}
H(y \ll 1)
\propto
y^{2/3}
\text{ and }
H(y \gg 1)
\propto y^{1/2}
\end{equation}

The non-Fermi-liquid scale is determined by setting
\begin{equation}
|\Sigma(T_{\rm NFL})|
\sim
T_{\rm NFL}.
\end{equation}
In terms of $y_{\rm NFL}
\equiv
T_{\rm NFL}/\Theta,$
this becomes the dimensionless relation
\begin{equation}
g H(y_{\rm NFL})
\sim
y_{\rm NFL}
\implies
g \sim \frac{y_{\rm NFL}}{H(y_{\rm NFL})}.
\end{equation}
%The value of $g$ therefore determines whether the NFL onset is encountered on the high-energy $z=2$ side or on the low-energy $z=3$ side of the crossover.
For $y_{\rm NFL} \gg 1$ we immediately get
\begin{equation}
    g \sim 
    \sqrt{y_{\rm NFL}} 
    \implies 
    T_{\rm NFL} \sim g^2 \Theta \gg \Theta
\end{equation}

This is the $z_{\rm eff}=2$ regime
and gives us the induced superconducting temperature (up to numerical pre-factors)
\begin{equation}
\label{eq:T_c_induced}
T_{\rm IND}
\sim
T_{\rm NFL}
\sim
\frac{M u^2n_b^2}
{v_F^2 \sqrt{\nu_F} (\nu_F - 4u n_b M C_F)^{3/2}}.
\end{equation}
On cooling from
higher $T$, the system therefore becomes non-Fermi-liquid
while it is still in the $z_{\rm eff}=2$ regime, and only subsequently
crosses into the $z=3$ regime at lower $T$.

The final round of self-consistency  requires $T_{\rm IND} \ll T_b$ which justifies the use of Eq.\eqref{eq:Low_T:Boson_Bubble_Inverse}.
The most transparent check is to simply plug in a parabolic $\varepsilon_{\textbf{k}} = k^2/2m_f - \mu_f$ which immediately gives us $\nu_F \sim m_f, \nu_F' = \nu_F'' = C_F = 0$ giving us
\begin{equation}
    \frac{T_{\rm IND}}{T_b} \sim \left(\frac{M}{m_f}\right)^2 \frac{n_b u^2}{v_F^2} \ll 1
\end{equation}

which can be easily satisfied by a weakly repulsive dilute gas.

It is easy to see that $T_{\rm IND}$ on its own isn't particularly impressive which takes us to the next part.

\section{Enhancing Superconductivity}
\label{sec:Enhancing}

If the original metal is already a superconductor then $V_{ff}(\bar{q})$ can be used to enhance $T_c$ irrespective of the pairing channel as was shown for the case for spherically symmetric Fermi surfaces in \cite{Metlitski_Cooper_Pairs}.
In this case it is reasonable to assume that $T_c \gg T_b$ and at this point the problem becomes much more complicated necessitating a closer look at all sorts of hair-splitting details \cite{Chubukov2020_interplay, Patel_Sachdev_Large_N, Thermal_Superconductivity}.

The very first thing that changes is the inverse boson polarization which for $\Omega_m = 0$ is now
\begin{equation}
\Pi_b^{-1}(q,0)
=
\frac{T}{n_b}
+
\frac{q^2}{12 Mn_b}
+\cdots ,
\end{equation}
whereas for $\Omega_m \neq 0$ it is
\begin{equation}
\Pi_b^{-1}(q,i\Omega_m)
=
\frac{M\Omega_m^2}{n_b q^2}
+
\frac{3T}{n_b}
+
\frac{q^2}{4M n_b}
\left(
1-\frac{24T^2}{\Omega_m^2}
\right)
+\cdots .
\end{equation}
This drastic change comes about because the discreet Matsubara frequency $\Omega_m \equiv 2\pi mT$ jumps quite strongly when going from $m = 0$ to $m=1$ for $T \gg T_b$ and the expansion has to be carried out differently for each case.
The derivations can be found in Appendix.[\ref{app:Bose_Lindhard}].
In the previous section we had taken quite some care in analyzing the interplay of $z=2$ vs $z=3$ criticality.
At higher $T$ we expect the system to be even more robustly described by $z=2$ physics when $\Omega_m^2/q^2$ is expected to dominate over $\Omega_m/q$ even more strongly.

%In the Eliashberg equations the $\Omega_m=0$ interaction corresponds to the term with equal incoming and outgoing fermionic Matsubara indices, $n=n'$.  This term enters the pairing vertex and fermion self energy in the same multiplicative manner and can therefore be removed by redefining the two Eliashberg functions.  The pairing instability is consequently controlled by the remaining $n\neq n'$ dynamical interaction, .

A closer look at the derivation of the pairing gap equations in \cite{Chubukov2020_interplay} shows that the $\Omega_m=0$ piece of $V_{ff}(\bar{q})$, which is controlled by $\Pi_b^{-1}(q, 0)$ corresponds to purely thermal contribution to the gap equations and can be absorbed by appropriate  re-definitions of the pairing gap and non-Fermi liquid self energy as reviewed briefly in Appendix.[\ref{app:gamma_model}].
The modified gap equations in terms of the re-defined quantities  captures the superconductivity in the system and are now controlled purely by $\Omega_m \neq 0$ part of $V_{ff}(\bar{q})$ which means we only have to care for $\Pi_b^{-1}(q,i\Omega_m)$.

The strongly $T$ dependent piece in $\Pi_b^{-1}(q,i\Omega_m)$ goes as
\begin{equation}
    \frac{24T^2}{\Omega_m^2} = \frac{6}{m^2\pi^2}
\end{equation}
which quickly decays to $0$ for $m \geq 1$ and can be safely neglected for qualitative analysis allowing us to write down the relevant effective fermion-fermion attraction
\begin{equation}
\label{eq:Effective_ff_High_T}
     V_{ff}(\bar{q}) 
    \approx
     \frac{-2u n_b/\nu_F}{6T + 2M\frac{\Omega_m^2}{q^2}  + c^2q^2}
\end{equation}
where we have neglected the $\Omega_m/q$ and $T^2/\Omega_m^2$ pieces as argued earlier which is basically the same as Eq.\eqref{eq:Effective_ff_Low_T} except for the additional $6T$ piece.

At this point there are two distinct serendipitous features that kick in simultaneously.

Firstly, a finite $T$ analysis of quantum critical metals inevitable runs into the problem of the boson thermal mass $\mathcal{M}(T)$ which are thermal corrections to the critical boson propagator \cite{IR_Div_Cure_Thermal_Mass, Thermal_Mass_Transport, Hertz, Millis, Sachdev_Phases}.
These effects were always ignored in any and all works on the $\gamma$ model \cite{Chubukov2020_interplay, GammaI, GammaII, GammaIII, GammaIV, GammaV, GammaVI} even for the case of finite $T$ so we cannot extrapolate any insights regarding this aspect from there.

In our work, $V_{ff}(\bar{q})$ is the direct analogue the critical boson propagator.
Now for $d=2$ it is known that $\mathcal{M}(T) \sim T$ up to logarithmic corrections and is independent of $z$  so the $6T$ in our $V_{ff}$ denominator has the same structural form which we inflate the vocabulary a bit and call the thermal boson mass of $V_{ff}$.
Note that the same reasoning doesn't apply for $d=3$ where $\mathcal{M}(T) \sim T^{1+1/z}$ \cite{Millis} which grows faster than $T$ in the denominator of $V_{ff}$.

%which basically comes from the static high $T$ compressibility of dilute bose gas which goes as $\Pi_b(0) \sim  n_b/T$ and has nothing to do with the dimensionality of the system and remains unmodified accordingly.

Secondly, as was discussed in \cite{Patel_Sachdev_Large_N}, $\mathcal{M}(T)$ is detrimental to critical superconductivity as it gives rise to a thermal contribution to fermion self energy $\Sigma_T(i\omega_n) \sim -i \operatorname{sgn}(\omega_n) \mathcal{X}\sqrt{T}$ (again neglecting log corrections) which for the case of $z=3$ ends up completely dominating the fermion self energy $\Sigma(i\omega_n) \propto |\omega_n|^{2/3}$ close to $T=0$ and at finite $T$ can wash away superconductivity if $\mathcal{X}$ is large where $\mathcal{X}$ depends on all sorts of system specific details.
For our case of $z=2$, $\Sigma_T(i\omega_n) \propto \sqrt{T}$ does \textit{not} change from as it doesn't depend on $z$ whereas $\Sigma(i\omega_n) \propto  \sqrt{|\omega_n|}$ 
as a consequence $\Sigma_T(i\omega_n)$ is now much less likely to dominate over $\Sigma(i\omega_n)$ and similarly much less likely to wash away critical superconductivity.

Having addressed all these technical and conceptual caveats, we are now in a position to analyze the case where the enhancement to $T_c$ is most viscerally transparent.
We couple a $z=2$ quantum critical superconductor\footnote{In 2D, physically realistic $z=2$ critical metals are found at SDW/CDW order \cite{Chubukov_Magnetic_AFM_Critical}. Starting from high $T$, the crucial enhancements in pairing takes place for electrons in the vicinity of the hot spots.}  with its own critical pairing interaction $V_C$ and already has $T_c \propto \mathfrak{g}$.

To see the enhancement in $T_c$ following \cite{Chubukov2020_interplay}, we need to combine the original dynamic interaction 
\begin{equation}
    V_C(z = 2,\Omega_m) 
    = 
    -\left(\frac{\mathfrak{g}}{\Omega_m} \right)^{1/2},
\end{equation} 
to our appropriately tuned excitonic gas with 
\begin{equation}
    V_{ff}(\Omega_m) \equiv \int \frac{dq}{v_F} \, V_{ff}(\bar{q}) = -\left(\frac{\mathfrak{g}'}{\Omega_m} \right)^{1/2}
\end{equation}
where we have plugged in Eq.\eqref{eq:Effective_ff_High_T} having neglected the "thermal mass" $6T$ piece in the denominator. 
This gives us
\begin{equation}
\label{eq:Enhancement_Final}
    \mathfrak{g}' 
    \sim\frac{M u^2 n_b^2/\sqrt{\nu_F}}{v_F^2\left(\nu_F- 4 u n_b M C_F\right)^{3 / 2}},
\end{equation}
which shows that $\mathfrak{g}' \sim T_{\rm IND}$ from the induced case in Eq.\eqref{eq:T_c_induced}.
Interestingly, this tells us that up to numerical pre-factors, $T_{\rm IND}$ actually doesn't depend whether we're below or above $T_b$. 

Now we can directly combine the two interactions to write down a total effective fermion attraction 
\begin{equation}
    V(\Omega_m) = -\left(\frac{\mathfrak{g}}{\Omega_m} \right)^{1/2}  - \left(\frac{\mathfrak{g}'}{\Omega_m} \right)^{1/2} 
    \equiv -\left(\frac{\mathfrak{g}''}{\Omega_m} \right)^{1/2}
\end{equation}
which gives us the main takeaway result in the form of enhanced 
\begin{equation}
    T_{\rm ENH} \propto \mathfrak{g}'' = \mathfrak{g} + \mathfrak{g}' + 2 \sqrt{\mathfrak{g} \mathfrak{g}'}.
\end{equation}

For any other $V_C(z \neq 2, \Omega_m)$ the extremely simple direct addition doesn't work and we have to go back to the original gap equations from \cite{Chubukov2020_interplay} to work out the $T_{\rm ENH}$ which cannot be given a simple analytical form.

As a side note, in \cite{Thermal_Boson_Paper}, tremendous amount of stress was made regarding the point that the enhancement to $T_c$ does not grow boundlessly with rising $M$ which seems to be in direct contradiction to Eq.\eqref{eq:Enhancement_Final}.
To this we only remark that throughout we have assumed that $T_c \ll T_F$ which allowed us to use all the patch approximations to perform the analysis and naïvely setting $M$ and therefore $T_c$ too large breaks this assumption down making the results completely unreliable.
The precise manner in which the simple formulas here break down at large $M$ cannot be answered within the scope of this analysis.

\section{Bottleneck}
\label{sec:Bottlenecks}

In what follows, $d$ refers to the distance between positive and negative charges in the bi-layer exciton and $L \gg d$ refers to the distance between the electron layer and exciton bi-layer in Fig.[\ref{fig:ATP}].
Till now we had been  treating the interaction strengths $u, \lambda$ as tunable numbers and focused our attention on $\Pi_b(\bar{q}), \Pi_f(\bar{q})$.
Now we discuss what happens when we pay a closer look at the interactions themselves.

%The one major issue we mentioned briefly earlier and postponed till now is with regards to the physical feasibility of the parameter tuning that leads us to criticality. 
The proper repulsive interaction experienced by the dipolar excitons is given by \cite{Exciton_Correlation}
\begin{equation}
u(r)=\frac{\mathcal{E}}{2\pi}
\left(
\frac{1}{r}
-
\frac{1}{\sqrt{r^2+d^2}}
\right).
\end{equation}
The Fourier transform in two dimensional momentum space can be carried out using the zeroth order Bessel's function (from hereon $q \equiv |\textbf{q}| > 0$)
\begin{equation}
\begin{aligned}
    u(q)
&\equiv
\int d^2r\,e^{-i\mathbf q\cdot\mathbf r} u(r)
=
\int_0^{\infty} dr \, r \, u(r) \int_0^{2 \pi} d \theta e^{-i q r \cos \theta}
\\
&=
\mathcal{E}
\int_0^\infty dr\,rJ_0(qr)
\left(
\frac{1}{r}
-
\frac{1}{\sqrt{r^2+d^2}}
\right).
\end{aligned}
\end{equation}
Using
\begin{equation}
\int_0^\infty dr\,J_0(qr)=\frac{1}{q},
\qquad
\int_0^\infty dr\,
\frac{rJ_0(qr)}{\sqrt{r^2+d^2}}
=
\frac{e^{-qd}}{q},
\end{equation}
we get ($u_0 \equiv \mathcal{E}d$)
\begin{equation}
\begin{aligned}
    u(q)
&=
\frac{\mathcal{E}}{q}
\left(1-e^{-qd}\right)
\\
&=
u_0
\left[
1-\frac{qd}{2}
+\frac{q^2d^2}{6}
+O(q^3d^3)
\right].
\end{aligned}
\end{equation}

This $-qd$ piece up above is genuinely problematic.
If we focus on the static denominator  of $V_{ff}(\bar{q})$ with this piece in place as well we get ($a, b, c$ are simple placeholders for clarity)
\begin{equation}
    V_{ff}^{-1}(\Omega = 0, q) \sim a-b q+c q^2,
\end{equation}
and we immediately see a minimum at $q_c= b/(2c)$ that gives us an effective interaction that is peaked at finite $q \neq 0$ which means the result from \cite{Metlitski_Cooper_Pairs} regarding channel independent enhancement becomes unusable.

This issue is circumvented by taking into account the $q$ dependence of the boson-fermion interaction as \cite{Lifesaver_Valley}
\begin{equation}
    \lambda(q)=\lambda_0 e^{-L q} \frac{1-e^{-qd}}{qd} = \lambda_0\left[1-\left(L+\frac{d}{2}\right)q+\cdots\right] ,
\end{equation}
then the criticality condition $u_0 = \nu_F \lambda_0^2$ immediately gives us
\begin{equation}
    u(q)-\nu_F \lambda^2(q)=u_0\left(2 L+\frac{d}{2}\right)q+O\left(q^2\right)>0
\end{equation}
circumventing the finite $q$ instability.

However this immediately modifies the effective fermion fermion interaction as
\begin{equation}
    V_{ff}^{-1}(\bar{q}) \sim v \,q + c^2q^2 +  \frac{\Omega^2}{q^2} \approx v \,q + \frac{\Omega^2}{q^2},
\end{equation}
where the final approximate expression is for high enough $T$ and small enough $q$, giving us $z=3/2$ for $\Omega \sim q^z$.

%With this resolution the qualitative aspects of enhanced critical superconductivity across all pairing channels for spherically symmetric Fermi surfaces is restored.
The resolution drastically changes the nature of criticality of the Fermi surface as now we have a marginal Fermi liquid (MFL) with $\gamma = 0$ as discussed in Sec.[\ref{subsec:Exponents}] and self energy $\Sigma(i\omega) \sim g |\omega| \ln(\Lambda/|\omega|)$ \cite{MFL}.

This gives
\begin{equation}
    T_{\rm MFL} \sim \Lambda e^{-1/g}
\end{equation}
as the crossover temperature to Fermi liquid physics which means that the parametric enhancements due to criticality are washed away.

%The parameter matching scheme from previous sections can be reapplied for this new case in much the same way but any deeper look into the enhancement of $T_c$ necessitates a full solution of the $\gamma$ model gap equations which is much beyond the scope of this work.

\section{Summary}
\label{sec:summary}

We coupled a two-dimensional metal to an uncondensed dilute bose gas and studied the resulting effective interactions within RPA. The induced fermion-fermion interaction
$V_{ff}(\bar q)=-\lambda^2[\Pi_b^{-1}(\bar q)+u-\lambda^2\Pi_f(\bar q)]^{-1}$
becomes long ranged on tuning to the mechanical instability,
$\lambda^2=u\nu_F^{-1}$, without simultaneously making the effective boson-boson and boson-fermion interactions singular.

If the $T_c$ of the original metal is extremely low then the bose gas essentially induces superconductivity at a temperature $T_{\rm IND}$.
For heavy dilute bosons the long ranged pairing is controlled by the bose gas fluctuations with $z=2$ 
and gives
\begin{equation}
T_{\rm IND}\sim T_{\rm NFL}
\sim
\frac{
M u^2 n_b^2
}{
v_F^2\sqrt{\nu_F}
\left(\nu_F-4u n_b M C_F\right)^{3/2}
},
\end{equation}
and we showed that it doesn't matter whether $T_{\rm IND}$ is above or below $T_b = 2\pi n_b/M$.

If the metal has a substantial $T_c$ of its own, then the bose gas enhances the superconducting temperature to $T_{\rm ENH}$. 
%Above $T_b$, the nonzero Matsubara sector acquires the additional thermal term $3T/n_b$. 
When the original metal is itself a $z=2$ quantum critical superconductor with $T_c \sim\mathfrak{g}$, the original and bose-gas-induced interactions have the same frequency dependence and their pairing scales combine directly,
\begin{equation}
T_{\rm ENH}
\sim
\mathfrak{g} + \mathfrak{g}^\prime+2\sqrt{\mathfrak{g}\mathfrak{g}^\prime},
\end{equation}
with $\mathfrak{g}' \sim T_{\rm IND}$.
The enhancement is therefore ultimately limited by the dilute boson density.

For a realistic electron-exciton mixtures \cite{Exciton_Correlation, Lifesaver_Valley}, the momentum dependence of $u(q)$ alone shifts the maximum of the induced attraction to finite
momentum. 
Including the corresponding momentum dependence of $\lambda(q)$ restores a forward-peaked interaction, but changes the low-energy form to $V_{ff}^{-1}(\bar q)\sim q+\Omega_m^2/q^2$ and hence to $\gamma=0$ in two dimensions. 
The fermions are then only marginally
non-Fermi liquid, $\Sigma(i\omega)\sim g|\omega|\ln(\Lambda/|\omega|)$, with the corresponding scale $T_{\rm MFL}\sim\Lambda e^{-1/g}$ \cite{MFL}. 
Thus the algebraic enhancement
associated with $\gamma>0$ quantum criticality is lost in the realistic electron-exciton realization.
This issue can only be resolved on engineering an interaction which decays slower in momentum space as compared to \cite{Exciton_Correlation, Lifesaver_Valley} which lies completely outside the scope of this work.

\section{Acknowledgments}
\label{sec:acknowledgments}

We thank Stefan Kehrein for introducing us to this topic and Sri Raghu for invaluable remarks as usual.

\renewcommand*{\bibfont}{\footnotesize}

\appendix

\section{The $\gamma$-model and quantum critical pairing}

\label{app:gamma_model}

We briefly summarize the $\gamma$-model framework used in the main text
\cite{Chubukov2020_interplay}.

The microscopic interaction between two fermions generally depends on both
the transferred momentum and frequency, $V(\mathbf q,i\Omega_m)$.  For the
low energy pairing problem both fermions are restricted to the vicinity of
the Fermi surface.  After projecting the interaction onto the pairing
channel of interest and carrying out the remaining momentum integration,
the Eliashberg equations can be written in terms of an effective interaction
which depends only on the transferred bosonic Matsubara frequency,
$\Omega_m=\omega_n-\omega_{n'}$.  We denote this frequency dependent pairing
kernel by $V(\Omega_m)$.  
For the forward scattering problem
of interest here,
\begin{equation}
V(\Omega_m)\equiv
\int \frac{dq_{\parallel}}{v_F}\,
V(q_{\parallel},i\Omega_m),
\end{equation}
where $q_{\parallel}$ is the momentum tangential to the local Fermi surface
patch. 
The $\gamma$-model describes the case where this momentum-integrated
interaction has the low frequency form
\begin{equation}
V(\Omega_m) = -\left(\frac{\mathfrak{g}}{|\Omega_m|}\right)^\gamma.
\end{equation}

where $\mathfrak{g}$ has dimensions of energy.  
The same interaction which produces
pairing also produces a non-Fermi liquid fermion self energy, so that the pairing problem has to be treated together with the fermion self energy
within Eliashberg theory.

At finite temperature the $\Omega_m=0$ contribution produces a singular
thermal piece.  
Following~\cite{Chubukov2020_interplay}, this contribution can be absorbed into redefined pairing and self-energy functions
$\Phi_m^*$ and $\widetilde{\Sigma}_m^*$.  
The remaining Eliashberg equations
are
\begin{equation}
\begin{aligned}
    \Phi_m^*
&=
\pi T \mathfrak{g}^\gamma
\sum_{m'\neq m}
\frac{\Phi_{m'}^*}
{\sqrt{(\widetilde{\Sigma}_{m'}^*)^2+(\Phi_{m'}^*)^2}}
\frac{1}{|\omega_m-\omega_{m'}|^\gamma},
\\
\widetilde{\Sigma}_m^*
&=
\omega_m+
\pi T \mathfrak{g}^\gamma
\sum_{m'\neq m}
\frac{\widetilde{\Sigma}_{m'}^*}
{\sqrt{(\widetilde{\Sigma}_{m'}^*)^2+(\Phi_{m'}^*)^2}}
\frac{1}{|\omega_m-\omega_{m'}|^\gamma}.
\end{aligned}
\end{equation}
with $\omega_m=(2m+1)\pi T$. 

The highest $T$ at which the equations have a non-zero eigenfunction defines $T_p$.  
Rescaling all frequencies by $\mathfrak{g}$ removes the microscopic scale from the equations and therefore gives $T_p = C_\gamma \mathfrak{g}$ where $C_\gamma$ is a dimensionless number determined by the gap equation.  
The normal-state self energy similarly behaves as $\Sigma(i\omega)\sim \mathfrak{g}^\gamma|\omega|^{1-\gamma}$, so that $T_{\mathrm{NFL}}\sim \mathfrak{g}$ and consequently $T_p\sim T_{\mathrm{NFL}}$ up
to numerical factors.  
This is the parametric relation used throughout
the main text.  
More precisely $T_p$ is the onset of pairing, while the true superconducting transition is lower but we do not distinguish these scales in our estimates of $T_c$.

The bose gas induced interaction studied here gives $\gamma = 1/2$. 
If the original critical interaction in the metal also has $\gamma = 1/2$ then we can directly combine the interactions to get,
\begin{equation}
\begin{aligned}
   V(\Omega_m)
&=
-\left(\frac{\mathfrak{g}}{|\Omega_m|}\right)^\gamma
-\left(\frac{\mathfrak{g}'}{|\Omega_m|}\right)^\gamma
=
-\left(\frac{\mathfrak{g}''}{|\Omega_m|}\right)^\gamma ,
\\
\mathfrak{g}''
&=
(\mathfrak{g}^\gamma + (\mathfrak{g}')^\gamma)^{1/\gamma}.
\end{aligned}
\end{equation}
Thus for
$\gamma=1/2$,
\begin{equation}
\mathfrak{g}''
=
\mathfrak{g}
+
\mathfrak{g}'
+
2\sqrt{\mathfrak{g} \mathfrak{g}'},
\qquad
T_c\sim T_p\sim \mathfrak{g}'',
\end{equation}
which gives the direct enhancement.  
If the two interactions have different $\gamma$'s, they cannot be combined into a single effective coupling and the full linearized Eliashberg
equation has to be solved.

\section{Long-wavelength fermionic Lindhard function for a general isotropic two-dimensional dispersion}
\label{app:fermion_lindhard_general}

We consider a two-dimensional fermionic system with an isotropic
single-particle dispersion $\varepsilon_{\mathbf{k}}$ and assume that the Fermi surface is smooth circle.
The polarization bubble of our interest is
\begin{equation}
\Pi_f(\mathbf{q},i\Omega_m)
=
\int\frac{d^2k}{(2\pi)^2}
\frac{
n_F(\varepsilon_{\mathbf{k}})
-
n_F(\varepsilon_{\mathbf{k}+\mathbf{q}})
}{
\varepsilon_{\mathbf{k}+\mathbf{q}}
-
\varepsilon_{\mathbf{k}}
-
i\Omega_m
}.
\label{eq:fermion_lindhard}
\end{equation}

\subsection{Local Fermi-surface quantities}

For general lattice systems, it is much more useful to express physical quantities of interest in terms of quantities that are  measurable and well defined in the vicinity of the Fermi surface which is what we do in this section.
For eg. the Fermi velocity $v_F$ is a local Fermi surface quantity whereas the Fermi momenta $k_F$ is not.

For a given fermion dispersion $\varepsilon_{\textbf{k}}$, the Fermi surface is determined implicitly by $\varepsilon_{\textbf{k}_F}=0$ with $\textbf{k}_F \equiv k_F \hat{n}$ where $k_F$is the angle independent is the Fermi momenta for an isotropic surface.
This yields us the two curvatures of the Fermi surface along  the normal $a(k)$ and tangential $\kappa(k)$ directions.  
With $k \equiv |\textbf{k}|$ we define
\begin{equation}
\begin{aligned}
    a(k)
&\equiv \varepsilon''_k =
\frac{d^2\varepsilon_k}{dk^2}
\implies 
a_F\equiv a(k_F)
\\
\kappa(k)
&\equiv
\frac{1}{k}\frac{d\varepsilon_k}{dk}
=
\frac{v(k)}{k}
\implies
\kappa_F\equiv \kappa(k_F).
\end{aligned}
\label{eq:a_definition}
\end{equation}
The Fermi velocity is defined by the radial velocity $v(k)\equiv \frac{d\varepsilon_k}{dk},$ right at the Fermi surface with $v_F \equiv v(k_F)$

A local dispersion expansion in the vicinity of a point $\textbf{k}_F$ on the Fermi surface takes the form
\begin{equation}
\varepsilon_{\mathbf{k}_F+\mathbf{q}}
=
v_Fq_\perp
+
\frac{a_F}{2}q_\perp^2
+
\frac{\kappa_F}{2}q_\parallel^2
+\cdots .
\label{eq:local_dispersion}
\end{equation}

The density of states is defined by
\begin{equation}
\nu(\epsilon)
\equiv
\int\frac{d^2k}{(2\pi)^2}
\delta\!\left(\epsilon-\varepsilon_{\mathbf{k}}\right)
=
\frac{1}{2\pi}
\int_0^\infty k\,dk\,
\delta\!\left(
\epsilon-\varepsilon_k
\right).
\label{eq:DOS_definition}
\end{equation}

Let $k(\epsilon)$ denote the momentum satisfying the implicit equation $\varepsilon_{k(\epsilon)}=\epsilon$, for eg. $k(\epsilon = 0) = k_F$ and define
\begin{equation}
v(\epsilon)
\equiv
\left.
\frac{d\varepsilon_k}{dk}
\right|_{k=k(\epsilon)}.
\end{equation}
For a locally monotonic dispersion with $v(\epsilon)>0$, the identity
\begin{equation}
\delta\!\left(
\epsilon-\varepsilon_{k(\epsilon)}
\right)
=
\frac{
\delta\!\left(k-k(\epsilon)\right)
}{
v(\epsilon)
}
\end{equation}
then gives (from the original definition of $\kappa$)
\begin{equation}
\nu(\epsilon)
=
\frac{1}{2\pi}
\frac{k(\epsilon)}{v(\epsilon)}
=
\frac{1}{2\pi \kappa(\epsilon)}.
\label{eq:DOS_isotropic}
\end{equation}

We use the following notation to define local quantities right at the Fermi surface
\begin{equation}
\begin{aligned}
    \nu_F
    &\equiv
    \nu(0),
    \\
    \nu_F'
&\equiv
\left.
\frac{d\nu(\epsilon)}{d\epsilon}
\right|_{\epsilon=0},
\\
\nu_F''
&\equiv
\left.
\frac{d^2\nu(\epsilon)}{d\epsilon^2}
\right|_{\epsilon=0}.
\end{aligned}
\label{eq:DOS_derivatives}
\end{equation}

The isotropic form of the density of states also relates the normal
curvature $a_F$ to the other local Fermi-surface quantities.  Since
\begin{equation}
\begin{aligned}
    \frac{dk}{d\epsilon}
    =
    \frac{1}{v},
    \quad
    \frac{dv}{d\epsilon}
=
\frac{a}{v},
\end{aligned}
\end{equation}
taking an energy derivative of Eq.~\eqref{eq:DOS_isotropic} gives
\begin{equation}
\frac{\nu'(\epsilon)}{\nu(\epsilon)}
=
\frac{1}{k(\epsilon)v(\epsilon)}
-
\frac{a(\epsilon)}{v^2(\epsilon)}.
\end{equation}
Using $\kappa(\epsilon)=v(\epsilon)/k(\epsilon)$, this becomes
\begin{equation}
a(\epsilon)
=
\kappa(\epsilon)
-
v^2(\epsilon)
\frac{\nu'(\epsilon)}{\nu(\epsilon)}
\implies
a_F
=
\kappa_F
-
v_F^2\frac{\nu_F'}{\nu_F}.
\label{eq:a_elimination}
\end{equation}
\label{eq:DOS_identity}

A second useful identity also follows directly from
$\kappa(\epsilon)=v(\epsilon)/k(\epsilon)$:
\begin{equation}
\frac{d\kappa}{d\epsilon}
=
-\kappa(\epsilon)
\frac{\nu'(\epsilon)}{\nu(\epsilon)}.
\label{eq:kappa_derivative}
\end{equation}

The final long-wavelength
polarization can therefore be expressed entirely in terms of the
local quantities
$v_F,
\kappa_F,
\nu_F,
\nu_F',
\nu_F'' ,$
all of which are in principle calculable directly from the original
dispersion $\varepsilon_{\textbf{k}}$ in an arbitrarily small neighborhood of
the Fermi surface.
In realistic systems, all these quantities become angle dependent.

\begin{widetext}

\subsection{Static expansion}

We first set $\Omega_m=0$ and choose $\mathbf{q} = q \hat{x}$  and set $\theta$ as
the angle between $\mathbf{k}$ and $\mathbf{q}$ to get
\begin{equation}
\Delta_{\mathbf{k},\mathbf{q}}
\equiv
\varepsilon_{\mathbf{k}+\mathbf{q}}
-
\varepsilon_{\mathbf{k}}
=
v(k)q\cos\theta
+
\frac{q^2}{2}
\left[
a(k)\cos^2\theta
+
\kappa(k)\sin^2\theta
\right]
+
O(q^3),
\label{eq:Delta_expansion_fermion}
\end{equation}

Expanding the numerator of Eq.~\eqref{eq:fermion_lindhard} gives
\begin{equation}
\frac{
n_F(\xi)
-
n_F(\xi+\Delta)
}{\Delta}
=
-n_F'(\xi)
-
\frac{\Delta}{2}n_F''(\xi)
-
\frac{\Delta^2}{6}n_F'''(\xi)
+\cdots .
\end{equation}
At $T=0$,
\begin{equation}
-n_F'(\xi)=\delta(\xi),
\qquad
-n_F''(\xi)=\delta'(\xi),
\qquad
-n_F'''(\xi)=\delta''(\xi),
\end{equation}
and hence
\begin{equation}
\begin{aligned}
    \Pi_f(q,0)
&=
\int\frac{d^2k}{(2\pi)^2}
\left[
\delta(\varepsilon_{\mathbf{k}})
+
\frac{\Delta_{\mathbf{k},\mathbf{q}}}{2}
\delta'(\varepsilon_{\mathbf{k}})
+
\frac{\Delta_{\mathbf{k},\mathbf{q}}^2}{6}
\delta''(\varepsilon_{\mathbf{k}})
+\cdots
\right]
\\
&= \nu_F + \int\frac{d^2k}{(2\pi)^2}
\left[
\frac{\Delta_{\mathbf{k},\mathbf{q}}}{2}
\delta'(\varepsilon_{\mathbf{k}})
+
\frac{\Delta_{\mathbf{k},\mathbf{q}}^2}{6}
\delta''(\varepsilon_{\mathbf{k}})
+\cdots
\right].
\end{aligned}
\label{eq:static_delta_expansion}
\end{equation}

Defining an angular average over the direction of $\mathbf{k}$ at fixed $k$ by
\begin{equation}
\left\langle
F(\theta)
\right\rangle_{\theta}
\equiv
\int_0^{2\pi}
\frac{d\theta}{2\pi}\,
F(\theta).
\label{eq:angular_average_definition}
\end{equation}
we get,
\begin{equation}
\begin{aligned}
    \left\langle
\cos\theta
\right\rangle_{\theta}
&=
\left\langle
\sin\theta
\right\rangle_{\theta}
=
\left\langle
\sin\theta\cos\theta
\right\rangle_{\theta}
=0,
\\
\left\langle
\cos^2\theta
\right\rangle_{\theta}
&=
\left\langle
\sin^2\theta
\right\rangle_{\theta}
=
\frac{1}{2},
\end{aligned}
\end{equation}
which gives
\begin{equation}
\Pi_f(q,0)
=
\nu_F+C_Fq^2+O(q^4),
\label{eq:static_general_form}
\end{equation}
where
\begin{equation}
C_F
=
-\frac{1}{8}
\left.
\frac{d}{d\epsilon}
\left[
\nu(\epsilon)
\left(
a(\epsilon)+\kappa(\epsilon)
\right)
\right]
\right|_{\epsilon=0}
+
\frac{1}{12}
\left.
\frac{d^2}{d\epsilon^2}
\left[
\nu(\epsilon)v^2(\epsilon)
\right]
\right|_{\epsilon=0}
=
\frac{\kappa_F\nu_F'}{12}
+
\frac{v_F^2}{24}
\left[
\nu_F''
-
2\frac{(\nu_F')^2}{\nu_F}
\right].
\label{eq:CF_intermediate}
\end{equation}

\subsection{Leading dynamical contribution}

We now look at $\Omega_m \neq 0$.  
At small $q$ we linearize the dispersion in the vicinity of the Fermi surface as $\Delta_{\mathbf{k},\mathbf{q}}
\simeq
v_Fq\cos\theta $ and $n_F(\varepsilon_{\mathbf{k}})
-
n_F(\varepsilon_{\mathbf{k}+\mathbf{q}})
\simeq
\delta(\varepsilon_{\mathbf{k}})
v_Fq\cos\theta.$

To leading order in the small $q$ expansion, Eq.~\eqref{eq:fermion_lindhard} therefore reduces to \cite{Chowdhury_Senthil}
\begin{equation}
\Pi_f(q,i\Omega_m)
\simeq
\nu_F
\int_0^{2\pi}
\frac{d\theta}{2\pi}
\frac{
v_Fq\cos\theta
}{
v_Fq\cos\theta-i\Omega_m
}
=
\nu_F
\left[
1-
\frac{|\Omega_m|}
{\sqrt{\Omega_m^2+v_F^2q^2}}
\right].
\label{eq:dynamical_angular}
\end{equation}

Combining this universal dynamical contribution with the analytic
static gradient expansion in the Landau-damped regime $|\Omega_m|\ll v_Fq,$ gives
\begin{equation}
\Pi_f(q,i\Omega_m)
=
\nu_F
+
C_Fq^2
-
\frac{\nu_F}{v_F}
\frac{|\Omega_m|}{q}
+\cdots.
\label{eq:PiF_complete}
\end{equation}

\section{Density bubble of a free two-dimensional Bose gas}

\label{app:Bose_Lindhard}

We consider a free two-dimensional Bose gas in the normal state,
\begin{equation}
\xi_{\mathbf{p}}
=
\frac{p^2}{2M}-\mu,
\qquad
\mu<0,
\end{equation}
and focus on the bosonic polarization bubble
\begin{equation}
\Pi_b(\mathbf{q},i\Omega_m)
=
 \int \frac{d^2 p}{(2 \pi)^2} \frac{n_B\left(\xi_{\textbf{p}+\textbf{q}}\right)-n_B\left(\xi_\textbf{p}\right)}{i \Omega_m-\xi_{\textbf{p}+\textbf{q}}+\xi_\textbf{p}},
\qquad
n_B(x)=\frac{1}{e^{x/T}-1}.
\label{eq:bose_lindhard_def}
\end{equation}

\subsection{Static Piece}

The static density bubble is simply the zero frequency dynamical bubble
\begin{equation}
\Pi_b(\mathbf{q},0)
=
-
\int\frac{d^2p}{(2\pi)^2}
\frac{
n_B(\xi_{\mathbf{p}+\mathbf{q}})
-
n_B(\xi_{\mathbf{p}})
}{
\xi_{\mathbf{p}+\mathbf{q}}
-
\xi_{\mathbf{p}}
}.
\end{equation}

At zero momentum,
\begin{equation}
\Pi_b(0,0)
=
-
\int\frac{d^2p}{(2\pi)^2}
n_B'(\xi_{\mathbf{p}})
=
\frac{\partial n_b}{\partial\mu}.
\end{equation}

We relate the bose gas density $n_b$ to the chemical potential using
\begin{equation}
n_b
=
\int\frac{d^2p}{(2\pi)^2}\,
n_B(\xi_{\mathbf{p}})
=
\int\frac{d^2p}{(2\pi)^2}\,
\frac{1}{e^{\xi_{\mathbf{p}}/T}-1}.
\end{equation}

In terms of the fugacity $z_b \equiv e^{\mu/T}$ and thermal wavelength $\lambda_T \equiv \sqrt{2\pi/(MT)}$ the boson density becomes
\begin{equation}
n_b
=
\frac{M}{2\pi}
\int_0^\infty d\epsilon\,
\frac{1}{e^{(\epsilon-\mu)/T}-1}
=
-\frac{1}{\lambda_T^2}\ln(1-z_b)
\implies
z_b
=
1-e^{-n_b\lambda_T^2}.
\end{equation}

Differentiating the density with respect to the chemical potential gives
\begin{equation}
\label{eq:Static_Bose_Bubble}
\Pi_b(0,0)
=
\frac{1}{T\lambda_T^2}
\frac{z_b}{1-z_b}
=
\frac{1}{T\lambda_T^2}
\left(
e^{n_b\lambda_T^2}-1
\right)
=
\frac{M}{2\pi}
\left[
\exp\left(
\frac{2\pi n_b}{MT}
\right)-1
\right]
=
\frac{M}{2\pi}
\left[
\exp\left(
\frac{T_b}{T}
\right)-1
\right].
\end{equation}
with $T_b \equiv 2\pi n_b/M$ being the bosonic quantum degeneracy temperature.

\subsection{Finite momentum and frequency dependence below degeneracy temperature}

We first consider the limit $T\ll T_b$
at fixed external momentum and frequency.
For convenience we define
\begin{equation}
\epsilon_q \equiv \frac{q^2}{2M},
\;
\text{ and }
\;
\eta_b \equiv n_b\lambda_T^2
=
\frac{2\pi n_b}{MT}
=
\frac{T_b}{T}.
\end{equation}
Eq.\eqref{eq:Static_Bose_Bubble} gives the zero-momentum static inverse
polarization
\begin{equation}
r_b(T)
\equiv
\Pi_b^{-1}(0,0)
=
\frac{2\pi}{M}
\frac{1}{e^{\eta_b}-1}.
\end{equation}

We consider $T\ll T_b$ which sends $\eta_b\rightarrow\infty$, and the chemical potential approaches zero exponentially,
\begin{equation}
\mu
=
T\ln\left(1-e^{-\eta_b}\right)
\simeq
-T e^{-\eta_b}.
\end{equation}

In the deeply degenerate two-dimensional ideal Bose gas the low-momentum occupation takes the form \cite{Hadzibabic2011}
\begin{equation}
n_B(\xi_p)\simeq
\frac{T}{\epsilon_p + |\mu|},
\end{equation}
with a characteristic momentum width
$p_c\sim\sqrt{2M|\mu|}$ which vanishes exponentially on cooling. 
Therefore, at low $T$ it peaks to an appropriately normalized Dirac delta function (this is a bit hand-wavy but can be made more rigorous following \cite{Hadzibabic2011}),
\begin{equation}
n_B(\xi_p)
\xrightarrow[T\to0]{}
(2\pi)^2n_b\,\delta^{(2)}(\mathbf p).
\end{equation}

Plugging this limit directly in the definition of $\Pi_b(\mathbf q,i\Omega_m)$ we obtain contributions from $\mathbf p=0$ and $\mathbf p=-\mathbf q$,
\begin{equation}
\Pi_b(\mathbf q,i\Omega_m)
\approx
n_b
\left[
\frac{1}{i\Omega_m+\epsilon_q}
-
\frac{1}{i\Omega_m-\epsilon_q}
\right]
=
\frac{2n_b\epsilon_q}
{\Omega_m^2+\epsilon_q^2}.
\end{equation}

The inverse polarization is therefore

\begin{equation}
\Pi_b^{-1}(\mathbf q,i\Omega_m)
\approx
\frac{q^2}{4Mn_b}
+
\frac{M\Omega_m^2}{n_bq^2}.
\end{equation}

\subsection{Finite momentum and frequency dependence above degeneracy temperature}
\label{app:bose_lindhard_highT}

Now we look at the nondegenerate regime $T\gg T_b$.
The fugacity now satisfies $z_b=e^{\mu/T}
\simeq
\frac{T_b}{T}
\ll 1,$
so the Bose distribution reduces to the Maxwell--Boltzmann
distribution $n_B(\xi_{\mathbf{p}})
\simeq
z_b e^{-p^2/(2MT)}
$

\subsubsection{Static sector: $\Omega_m=0$}

For zero Matsubara frequency, Eq.~\eqref{eq:bose_lindhard_def} becomes
\begin{equation}
\Pi_b(q,0)
=
\int\frac{d^2p}{(2\pi)^2}
n_B(\xi_{\mathbf{p}})
\frac{
1-e^{-\Delta_{\mathbf{p},\mathbf{q}}/T}
}{
\Delta_{\mathbf{p},\mathbf{q}}
}.
\end{equation}
with $\Delta_{\mathbf{p},\mathbf{q}}
\equiv
\xi_{\mathbf{p}+\mathbf{q}}
-
\xi_{\mathbf{p}}
=
\frac{\mathbf{p}\cdot\mathbf{q}}{M}
+
\frac{q^2}{2M}.$

Small $q$ expansion gives
\begin{equation}
\frac{1-e^{-\Delta/T}}{\Delta}
=
\frac{1}{T}
-
\frac{\Delta}{2T^2}
+
\frac{\Delta^2}{6T^3}
+\cdots .
\end{equation}

It is convenient to introduce a thermal average with respect to the
normalized Maxwell--Boltzmann distribution,
\begin{equation}
\left\langle
F(\mathbf{p})
\right\rangle_{\mathrm{MB}}
\equiv
\frac{1}{n_b}
\int\frac{d^2p}{(2\pi)^2}
n_B(\xi_{\mathbf{p}})
F(\mathbf{p}),
\label{eq:MB_average_definition}
\end{equation}

The basic momentum averages are
\begin{equation}
\left\langle 1\right\rangle_{\mathrm{MB}}=1 
\qquad\left\langle p_i\right\rangle_{\mathrm{MB}}=0,
\qquad \left\langle p_i p_j\right\rangle_{\mathrm{MB}}
=
MT\,\delta_{ij}
\implies\left\langle
\Delta_{\mathbf{p},\mathbf{q}}
\right\rangle_{\mathrm{MB}}
=
\frac{q^2}{2M},
\label{eq:MB_second_moment}
\end{equation}
and to the order required,
\begin{equation}
\left\langle
\Delta_{\mathbf{p},\mathbf{q}}^2
\right\rangle_{\mathrm{MB}}
=
\frac{q^2}{M^2}
\left\langle
(\mathbf{p}\cdot\hat{\mathbf{q}})^2
\right\rangle_{\mathrm{MB}}
+
O(q^4)
=
\frac{Tq^2}{M}
+
O(q^4).
\end{equation}

we therefore obtain
\begin{equation}
\Pi_b(q,0)
=
\frac{n_b}{T}
\left[
1-\frac{q^2}{12MT}
+
O\left(\frac{q^4}{M^2T^2}\right)
\right]
\implies
\Pi_b^{-1}(q,0)
=
\frac{T}{n_b}
+
\frac{q^2}{12Mn_b}
+
O(q^4).
\label{eq:PiB_static}
\end{equation}

\subsubsection{Nonzero Matsubara sector: $\Omega_m\neq0$}

For a fixed nonzero bosonic Matsubara frequency, $\Omega_m=2\pi m T,$ the small-$q$ expansion follows a qualitatively different route.
Shifting the momentum in the second term of
Eq.~\eqref{eq:bose_lindhard_def}, we can write
\begin{equation}
\Pi_b(q,i\Omega_m)
=
\int\frac{d^2p}{(2\pi)^2}
n_B(\xi_{\mathbf{p}})
\left[
\frac{1}{
\frac{\mathbf{p}\cdot\mathbf{q}}{M}
+\frac{q^2}{2M}
-i\Omega_m
}
-
\frac{1}{
\frac{\mathbf{p}\cdot\mathbf{q}}{M}
-\frac{q^2}{2M}
-i\Omega_m
}
\right].
\label{eq:Pi_nonzero_shifted}
\end{equation}

Writing the momentum component parallel to the external momentum as
$p_{\parallel}\equiv \mathbf{p}\cdot\hat{\mathbf{q}}$, it is convenient to define
\begin{equation}
x\equiv \frac{p_{\parallel}q}{M},
\qquad
y\equiv \frac{q^2}{2M}.
\end{equation}
The integrand can then be expanded as
\begin{equation}
\frac{1}{x+y-i\Omega_m}
-
\frac{1}{x-y-i\Omega_m}
=
\frac{2y}{\Omega_m^2}
-\frac{4ixy}{\Omega_m^3}
-\frac{6x^2y}{\Omega_m^4}
+\frac{8ix^3y}{\Omega_m^5}
+\frac{10x^4y}{\Omega_m^6}
-\frac{2y^3}{\Omega_m^4}
+O(q^8).
\end{equation}
The terms odd in $x$, and hence odd in $p_{\parallel}$, vanish upon
integration. 
The remaining Maxwell--Boltzmann averages are
\begin{equation}
\left\langle p_{\parallel}^2\right\rangle_{\rm MB}=MT,
\qquad
\left\langle p_{\parallel}^4\right\rangle_{\rm MB}
=3(MT)^2,
\qquad
\implies
\qquad
\left\langle x^2\right\rangle_{\rm MB}
=
\frac{Tq^2}{M},
\qquad
\left\langle x^4\right\rangle_{\rm MB}
=
\frac{3T^2q^4}{M^2}.
\end{equation}
Using $\int d^2p\,n_B(\xi_p)= 4\pi^2 n_b$, we now get
\begin{equation}
\Pi_b(q,i\Omega_m)
=
n_b
\left[
\frac{2y}{\Omega_m^2}
-\frac{6y}{\Omega_m^4}
\left\langle x^2\right\rangle_{\rm MB}
+\frac{10y}{\Omega_m^6}
\left\langle x^4\right\rangle_{\rm MB}
-\frac{2y^3}{\Omega_m^4}
+O(q^8)
\right].
\end{equation}
Substituting $y$ then gives
\begin{equation}
\Pi_b(q,i\Omega_m)
=
\frac{n_b q^2}{M\Omega_m^2}
-
\frac{3 n_b Tq^4}{M^2\Omega_m^4}
+
\frac{n_b q^6}{M^3}
\left[
\frac{15T^2}{\Omega_m^6}
-
\frac{1}{4\Omega_m^4}
\right]
+
O(q^8).
\label{eq:Pi_nonzero_expansion}
\end{equation}

Inverting
Eq.~\eqref{eq:Pi_nonzero_expansion} gives
\begin{equation}
\Pi_b^{-1}(q,i\Omega_m)
=
\frac{M\Omega_m^2}{n_b q^2}
+
\frac{3T}{n_b}
+
\frac{q^2}{Mn_b}
\left[
\frac{1}{4}
-
\frac{6T^2}{\Omega_m^2}
\right]
+
O(q^4).
\label{eq:Pi_inv_nonzero}
\end{equation}

\section{Boson self energy in the non-Fermi liquid regime}
\label{app:Boson_Self_Energy}
An immediate question is whether the onset of non-Fermi liquid behavior in the fermion sector feeds back strongly onto the bose gas. 
We address this using the standard $T$-matrix approximation for the boson self energy \cite{Extended_T_Matrix, Polaron_Equations, Polarons_Review, Demler_Polarons}. 
The free boson and dressed fermion propagators are
\begin{equation}
G_b^0(\bar p)
=
\frac{1}{i\nu_m-\xi_p},
\qquad
G_f(\bar k)
=
\frac{1}{W(i\omega_n)-\varepsilon_k},
\qquad
W(i\omega_n)\equiv i\omega_n-\Sigma(i\omega_n).
\end{equation}

The boson self energy in the ladder approximation defined in terms of the bose-fermi $T$-matrix $\Gamma_{bf}$ which in turn is defined via the the bose-fermi particle-particle bubble $\Pi_{bf}^{pp}$ as
\begin{equation}
\Sigma_b(\bar p)
=
\int_{\bar k}
G_f(\bar k)\,
\Gamma_{bf}(\bar p+\bar k),
\quad
\Gamma_{bf}^{-1}(\bar P)
=
\lambda^{-1}
-
\Pi_{bf}^{pp}(\bar P),
\quad
\Pi_{bf}^{pp}(\bar P)
=
\int_{\bar q}
G_b^0(\bar q)\,
G_f(\bar P-\bar q).
\end{equation}

Here $\bar P=(P,\omega_n)$ carries a fermionic Matsubara frequency.

We now ask whether $\Pi_{bf}^{pp}$ changes drastically when the fermions become incoherent.
To address this question we use the standard tricks from \cite{Sachdev_Phases} which is relevant as $\Pi_{bf}^{pp}$, with its one boson and one fermion line is closely related to the critical surface fermion self energy and the early part of the calculation can be borrowed directly.
%For the non-Fermi liquid regime of interest, $\Sigma(i\omega_n)$ dominates over the bare $i\omega_n$ term at sufficiently small frequency.

We consider the total pair momentum $P$ close to the Fermi surface and decompose the internal momentum into normal $q_\perp$ and tangential $q_\parallel$ components to expand the dispersions as
\begin{equation}
\varepsilon_{P-q}
\simeq
v_F(P_\perp-q_\perp)
+
\frac{\kappa_F}{2}(P_\parallel-q_\parallel)^2,
\qquad
\xi_q
=
\frac{q_\perp^2+q_\parallel^2}{2M}
-\mu.
\end{equation}

The critical Fermi-surface patch theory has the anisotropic
low-energy scaling $v_F q_\perp \sim \kappa_F q_\parallel^2$ \cite{Mross_Senthil,  Sachdev_QPT, Sachdev_Phases} and
consequently $q_\perp^2 \sim q_\parallel^4 \ll q_\parallel^2$ that feeds into the bose gas propagator which can now be approximated as
\begin{equation}
[G_b^0(\bar q)]^{-1}
=
i\nu_m - \frac{q_\perp^2 + q_\parallel^2}{2M}+\mu
\simeq
i\nu_m - \frac{q_\parallel^2}{2M}+\mu.
\end{equation}

The particle-particle bubble therefore becomes
\begin{equation}
\Pi_{bf}^{pp}(\bar P)
=
T\sum_{\nu_m}
\int\frac{dq_\perp dq_\parallel}{(2\pi)^2}
\frac{1}
{i\nu_m-\frac{q_\parallel^2}{2M}+\mu}
\frac{1}
{
W(i\omega_n - i\nu_m)
-v_F(P_\perp-q_\perp)
-\frac{\kappa_F}{2}(P_\parallel-q_\parallel)^2
}.
\end{equation}

Using the necessary condition \cite{Sachdev_Phases} $\mathrm{sgn}
\left[
\omega_n
-
\frac{\Sigma(i\omega_n)}{i}
\right]
=
\mathrm{sgn}(\omega_n),$
we obtain
\begin{equation}
\int\frac{dq_\perp}{2\pi}
G(\bar P-\bar q)
=
-\frac{i}{2v_F}
\mathrm{sgn}(\omega_n-\nu_m).
\end{equation}
Thus all explicit dependence on the non-Fermi liquid $\Sigma(i\omega)$ drops out and
\begin{equation}
\Pi_{bf}^{pp}(\bar P)
=
-\frac{i}{2v_F}
T\sum_{\nu_m}
\int\frac{dq_\parallel}{2\pi}
\frac{
\mathrm{sgn}(\omega_n-\nu_m)
}
{
i\nu_m-q_\parallel^2/(2M)+\mu
}.
\end{equation}
This is precisely the same expression obtained using a free Fermi gas propagator. 
Therefore, to leading order in the patch expansion, this indifference to $\Sigma(i\omega)$ climbs up the ladder $\Pi_{bf}^{pp,\mathrm{NFL}}(\bar P)
\simeq
\Pi_{bf}^{pp,\mathrm{FL}}(\bar P),$ which then immediately carries over to the $T$-matrix as $\Gamma_{bf}^{\mathrm{NFL}}(\bar P)
\simeq
\Gamma_{bf}^{\mathrm{FL}}(\bar P).$
Furthermore, the leading $T$-matrix obtained above is independent of the normal component of the pair momentum.
The normal momentum integral in the boson self energy can therefore again be carried out
\begin{equation}
\int\frac{dk_\perp}{2\pi}
G(\bar k)
=
-\frac{i}{2v_F}
\mathrm{sgn}(\omega_n)
\implies
\Sigma_b(\bar p)
=
-\frac{i}{2v_F}
T\sum_{\omega_n}
\int\frac{dk_\parallel}{2\pi}
\mathrm{sgn}(\omega_n)
\Gamma_{bf}(\bar p+\bar k).
\end{equation}

Consequently, $\Sigma_b^{\mathrm{NFL}}(\bar p)
\simeq
\Sigma_b^{\mathrm{FL}}(\bar p)$
which means that the bosons don't really care that much if the Fermi surface is critical or not.

\end{widetext}

% The \nocite command causes all entries in a bibliography to be printed out
% whether or not they are actually referenced in the text. This is appropriate
% for the sample file to show the different styles of references, but authors
% most likely will not want to use it.
%\nocite{*}

%\bibliography{apssamp}% Produces the bibliography via BibTeX.
%\bibliographystyle{abbrv}
\bibliography{main.bib}

@article{Hofmann2022Heuristic,
  author    = {Hofmann, Johannes S. and Chowdhury, Debanjan and Kivelson, Steven A. and Berg, Erez},
  title     = {Heuristic bounds on superconductivity and how to exceed them},
  journal   = {npj Quantum Materials},
  volume    = {7},
  pages     = {83},
  year      = {2022},
  doi       = {10.1038/s41535-022-00491-1},
  url       = {https://doi.org/10.1038/s41535-022-00491-1}
}

@article{Esterlis_Bounds,
  author    = {Esterlis, I. and Kivelson, S. A. and Scalapino, D. J.},
  title     = {A bound on the superconducting transition temperature},
  journal   = {npj Quantum Materials},
  volume    = {3},
  pages     = {59},
  year      = {2018},
  publisher = {Nature Publishing Group},
  doi       = {10.1038/s41535-018-0133-0}
}

@article{Resonantly_Enhanced,
  title     = {Theory of resonantly enhanced photo-induced superconductivity},
  author    = {Eckhardt, CJ and Chattopadhyay, A and Cavalleri, A and Demler, E and Sentef, MA},
  journal   = {Nature Communications},
  volume    = {15},
  pages     = {2291},
  year      = {2024},
  doi       = {10.1038/s41467-024-46632-x},
  url       = {https://www.nature.com/articles/s41467-024-46632-x}
}

@article{Transient_Pumped,
  title     = {Transient superconductivity from electronic squeezing of optically pumped phonons},
  author    = {Kennes, D. M. and Wilner, E. Y. and Reichman, D. R. and Millis, A. J.},
  journal   = {Nature Physics},
  volume    = {13},
  number    = {5},
  pages     = {479--483},
  year      = {2017},
  publisher = {Nature Publishing Group},
  doi       = {10.1038/nphys4024},
  url       = {https://arxiv.org/abs/1609.03802}
}

@misc{Thermal_Boson_Paper,
      title={Enhancing superconductivity using thermal bosons}, 
      author={Ekaterina Vlasiuk and Manfred Salmhofer and Eugene Demler and Richard Schmidt},
      year={2026},
      eprint={2603.06796},
      archivePrefix={arXiv},
      primaryClass={cond-mat.mes-hall},
      url={https://arxiv.org/abs/2603.06796},
}

@article{Chubukov2020_interplay,
  title = {The interplay between superconductivity and non-Fermi liquid at a quantum-critical point in a metal},
  author = {Chubukov, Andrey V. and Abanov, Artem and Wang, Yuxuan and Wu, Yi-Ming},
  journal = {Annals of Physics},
  volume = {417},
  pages = {168142},
  year = {2020},
  issn = {0003-4916},
  doi = {10.1016/j.aop.2020.168142},
  url = {https://doi.org}
}

@book{coleman2015introduction,
  title={Introduction to Many-body Physics},
  author={Coleman, Piers},
  year={2015},
  publisher={Cambridge University Press}
}

@book{Sachdev_Phases, 
place={Cambridge}, 
title={Quantum Phases of Matter}, 
publisher={Cambridge University Press}, 
author={Sachdev, Subir}, 
year={2023}}

@book{Sachdev_QPT, 
place={Cambridge}, 
edition={2}, 
title={Quantum Phase Transitions}, 
publisher={Cambridge University Press}, 
author={Sachdev, Subir}, 
year={2011}}

@book{bruus_Flensberg,
  title={Many-body Quantum theory in Condensed Matter Physics},
  author={Bruus, Henrik and Flensberg, Karsten},
  year={2004},
  publisher={Oxford university press}
}

@article{Sigrist2005Introduction,
  author  = {Sigrist, Manfred},
  title   = {Introduction to Unconventional Superconductivity},
  journal = {AIP Conference Proceedings},
  volume  = {789},
  pages   = {165--243},
  year    = {2005},
  doi     = {10.1063/1.2080350}
}

@book{mahan2000many,
  title     = {Many-Particle Physics},
  author    = {Mahan, Gerald D.},
  edition   = {3rd},
  year      = {2000},
  publisher = {Kluwer Academic/Plenum Publishers},
  address   = {New York},
  isbn      = {0-306-46338-5}
}

@article{Metlitski_Cooper_Pairs,
  title = {Cooper pairing in non-Fermi liquids},
  author = {Metlitski, Max A. and Mross, David F. and Sachdev, Subir and Senthil, T.},
  journal = {Phys. Rev. B},
  volume = {91},
  issue = {11},
  pages = {115111},
  numpages = {18},
  year = {2015},
  month = {Mar},
  publisher = {American Physical Society},
  doi = {10.1103/PhysRevB.91.115111},
  url = {https://link.aps.org/doi/10.1103/PhysRevB.91.115111}
}

@article{Exciton_Superconductor,
  title = {Superconductivity Induced by Strong Electron-Exciton Coupling in Doped Atomically Thin Semiconductor Heterostructures},
  author = {von Milczewski, Jonas and Chen, Xin and Imamoglu, Atac and Schmidt, Richard},
  journal = {Phys. Rev. Lett.},
  volume = {133},
  issue = {22},
  pages = {226903},
  numpages = {9},
  year = {2024},
  month = {Nov},
  publisher = {American Physical Society},
  doi = {10.1103/PhysRevLett.133.226903},
  url = {https://link.aps.org/doi/10.1103/PhysRevLett.133.226903}
}

@article{Near_Field_Coulomb_Drag,
  title = {Near-field heat transfer and drag resistance in bilayers of composite fermions},
  author = {Zverevich, Dmitry and Levchenko, Alex},
  journal = {Phys. Rev. B},
  volume = {109},
  issue = {15},
  pages = {155303},
  numpages = {11},
  year = {2024},
  month = {Apr},
  publisher = {American Physical Society},
  doi = {10.1103/PhysRevB.109.155303},
  url = {https://link.aps.org/doi/10.1103/PhysRevB.109.155303}
}

@article{Coulomb_Drag,
  title = {Coulomb drag},
  author = {Narozhny, B. N. and Levchenko, A.},
  journal = {Rev. Mod. Phys.},
  volume = {88},
  issue = {2},
  pages = {025003},
  numpages = {55},
  year = {2016},
  month = {May},
  publisher = {American Physical Society},
  doi = {10.1103/RevModPhys.88.025003},
  url = {https://link.aps.org/doi/10.1103/RevModPhys.88.025003}
}

@article{Coulomb_Drag_Graphene,
  title = {Coulomb drag in graphene--boron nitride heterostructures: Effect of virtual phonon exchange},
  author = {Amorim, Bruno and Schiefele, J\"urgen and Sols, Fernando and Guinea, Francisco},
  journal = {Phys. Rev. B},
  volume = {86},
  issue = {12},
  pages = {125448},
  numpages = {7},
  year = {2012},
  month = {Sep},
  publisher = {American Physical Society},
  doi = {10.1103/PhysRevB.86.125448},
  url = {https://link.aps.org/doi/10.1103/PhysRevB.86.125448}
}

@article{Collisionless_Sound,
  title = {Collisionless Sound in a Uniform Two-Dimensional Bose Gas},
  author = {Ota, Miki and Larcher, Fabrizio and Dalfovo, Franco and Pitaevskii, Lev and Proukakis, Nick P. and Stringari, Sandro},
  journal = {Phys. Rev. Lett.},
  volume = {121},
  issue = {14},
  pages = {145302},
  numpages = {5},
  year = {2018},
  month = {Oct},
  publisher = {American Physical Society},
  doi = {10.1103/PhysRevLett.121.145302},
  url = {https://link.aps.org/doi/10.1103/PhysRevLett.121.145302}
}

@article{Mihaila_Lindhard,
    author    = {Bogdan Mihaila},
    title     = {Lindhard function of a d-dimensional Fermi gas},
    journal   = {arXiv preprint arXiv:1111.5337},
    year      = {2011},
    eprint    = {1111.5337},
    archivePrefix = {arXiv},
    primaryClass  = {cond-mat.quant-gas}
}

@article{Fu_Kohn_Luttinger,
  title = {Nematic, chiral, and topological superconductivity in twisted transition metal dichalcogenides},
  author = {Schrade, Constantin and Fu, Liang},
  journal = {Phys. Rev. B},
  volume = {110},
  issue = {3},
  pages = {035143},
  numpages = {6},
  year = {2024},
  month = {Jul},
  publisher = {American Physical Society},
  doi = {10.1103/PhysRevB.110.035143},
  url = {https://link.aps.org/doi/10.1103/PhysRevB.110.035143}
}

@article{Kohn_Luttinger_Robustness,
	title={Robustness of the Kohn-Luttinger mechanism against symmetry breaking},
	volume={9},
	ISSN={2666-6812},
	url={http://doi.org},
	DOI={10.21468/scipostphyscore.9.2.034},
	number={2},
	journal={SciPost Physics Core},
	publisher={SciPost},
	author={Dalal, Amir and Ruhman, Jonathan and Kozii, Vladyslav},
	year={2026},
	month={jun}
}

@article{Collective_Modes,
  title = {Collective modes in a dilute Bose-Fermi mixture},
  author = {Yip, S. K.},
  journal = {Phys. Rev. A},
  volume = {64},
  issue = {2},
  pages = {023609},
  numpages = {4},
  year = {2001},
  month = {Jul},
  publisher = {American Physical Society},
  doi = {10.1103/PhysRevA.64.023609},
  url = {https://link.aps.org/doi/10.1103/PhysRevA.64.023609}
}

@article{Patel_Sachdev_Large_N,
  title = {Large-$N$ theory of critical Fermi surfaces},
  author = {Esterlis, Ilya and Guo, Haoyu and Patel, Aavishkar A. and Sachdev, Subir},
  journal = {Phys. Rev. B},
  volume = {103},
  issue = {23},
  pages = {235129},
  numpages = {32},
  year = {2021},
  month = {Jun},
  publisher = {American Physical Society},
  doi = {10.1103/PhysRevB.103.235129},
  url = {https://link.aps.org/doi/10.1103/PhysRevB.103.235129}
}

@article{Marsiglio_Eliashberg,
  title     = {Eliashberg theory: A short review},
  author    = {Marsiglio, Frank},
  journal   = {Annals of Physics},
  volume    = {417},
  pages     = {168102},
  year      = {2020},
  doi       = {10.1016/j.aop.2020.168102},
  url       = {https://doi.org},
  publisher = {Elsevier}
}

@article{IR_Div_Cure_Thermal_Mass,
  title = {How non-Fermi liquids cure their infrared divergences},
  author = {Damia, Jeremias Aguilera and Sol\'{\i}s, Mario and Torroba, Gonzalo},
  journal = {Phys. Rev. B},
  volume = {102},
  issue = {4},
  pages = {045147},
  numpages = {14},
  year = {2020},
  month = {Jul},
  publisher = {American Physical Society},
  doi = {10.1103/PhysRevB.102.045147},
  url = {https://link.aps.org/doi/10.1103/PhysRevB.102.045147}
}

@article{Thermal_Mass_Transport,
  title = {Transport near the Ising-nematic quantum critical point of metals in two dimensions},
  author = {Hartnoll, Sean A. and Mahajan, Raghu and Punk, Matthias and Sachdev, Subir},
  journal = {Phys. Rev. B},
  volume = {89},
  issue = {15},
  pages = {155130},
  numpages = {15},
  year = {2014},
  month = {Apr},
  publisher = {American Physical Society},
  doi = {10.1103/PhysRevB.89.155130},
  url = {https://link.aps.org/doi/10.1103/PhysRevB.89.155130}
}

@article{Hertz,
  title = {Quantum critical phenomena},
  author = {Hertz, John A.},
  journal = {Phys. Rev. B},
  volume = {14},
  issue = {3},
  pages = {1165--1184},
  numpages = {0},
  year = {1976},
  month = {Aug},
  publisher = {American Physical Society},
  doi = {10.1103/PhysRevB.14.1165},
  url = {https://link.aps.org/doi/10.1103/PhysRevB.14.1165}
}

@article{Millis,
  title = {Effect of a nonzero temperature on quantum critical points in itinerant fermion systems},
  author = {Millis, A. J.},
  journal = {Phys. Rev. B},
  volume = {48},
  issue = {10},
  pages = {7183--7196},
  numpages = {0},
  year = {1993},
  month = {Sep},
  publisher = {American Physical Society},
  doi = {10.1103/PhysRevB.48.7183},
  url = {https://link.aps.org/doi/10.1103/PhysRevB.48.7183}
}

@article{Esterlis_Critical_Eliashberg,
  author = {Esterlis, Ilya and Schmalian, J{\"o}rg},
  title = {Quantum Critical Eliashberg Theory},
  journal = {Annual Review of Condensed Matter Physics},
  year = {2026},
  volume = {17},
  number = {1},
  pages = {419--448},
  doi = {10.1146/annurev-conmatphys-032822-042856},
  url = {https://www.annualreviews.org/content/journals/10.1146/annurev-conmatphys-032822-042856}
}

@article{Phonon_Exchange,
  title = {Phonon exchange in dilute Fermi-Bose mixtures: Tailoring the Fermi-Fermi interaction},
  author = {Bijlsma, M. J. and Heringa, B. A. and Stoof, H. T. C.},
  journal = {Phys. Rev. A},
  volume = {61},
  issue = {5},
  pages = {053601},
  numpages = {11},
  year = {2000},
  month = {Apr},
  publisher = {American Physical Society},
  doi = {10.1103/PhysRevA.61.053601},
  url = {https://link.aps.org/doi/10.1103/PhysRevA.61.053601}
}

@article{Mechanical_instability,
  title = {Zero-temperature phase diagram of binary boson-fermion mixtures},
  author = {Viverit, L. and Pethick, C. J. and Smith, H.},
  journal = {Phys. Rev. A},
  volume = {61},
  issue = {5},
  pages = {053605},
  numpages = {8},
  year = {2000},
  month = {Apr},
  publisher = {American Physical Society},
  doi = {10.1103/PhysRevA.61.053605},
  url = {https://link.aps.org/doi/10.1103/PhysRevA.61.053605}
}

@article{Thermal_Superconductivity,
  title = {Thermal effects in non-Fermi liquid superconductivity},
  author = {Damia, Jeremias Aguilera and Sol\'{\i}s, Mario and Torroba, Gonzalo},
  journal = {Phys. Rev. B},
  volume = {103},
  issue = {15},
  pages = {155161},
  numpages = {21},
  year = {2021},
  month = {Apr},
  publisher = {American Physical Society},
  doi = {10.1103/PhysRevB.103.155161},
  url = {https://link.aps.org/doi/10.1103/PhysRevB.103.155161}
}

@article{2D_Bose_Gas_Normal_State,
  author = {Pieri, Pierbiagio and Strinati, G. C. and Tifrea, Ionel},
  title = {Two-dimensional dilute Bose gas in the normal phase},
  journal = {The European Physical Journal B},
  year = {2001},
  volume = {22},
  number = {1},
  pages = {79--91},
  doi = {10.1007/BF01325463},
  url = {https://link.springer.com/article/10.1007/BF01325463}
}

@article{Exciton_Correlation,
  title = {Exciton correlations in coupled quantum wells and their luminescence blue shift},
  author = {Laikhtman, B. and Rapaport, R.},
  journal = {Phys. Rev. B},
  volume = {80},
  issue = {19},
  pages = {195313},
  numpages = {12},
  year = {2009},
  month = {Nov},
  publisher = {American Physical Society},
  doi = {10.1103/PhysRevB.80.195313},
  url = {https://link.aps.org/doi/10.1103/PhysRevB.80.195313}
}

@article{Lifesaver_Valley,
  author    = {Boev, M. V. and Snegirev, A. V.},
  title     = {Valley Hall Effect in a Two-Dimensional Electron--Exciton System},
  journal   = {JETP Letters},
  volume    = {121},
  number    = {2},
  pages     = {103--110},
  year      = {2025},
  doi       = {10.1134/S0021364024604238},
  url       = {https://springer.com}
}

@article{Polarons_Review,
	doi = {10.1088/0034-4885/77/3/034401},
	url = {https://doi.org},
	year = 2014,
	month = {feb},
	publisher = {{IOP} Publishing},
	volume = {77},
	number = {3},
	pages = {034401},
	author = {Pietro Massignan and Matteo Zaccanti and Georg M Bruun},
	title = {Polarons, dressed molecules and itinerant ferromagnetism in ultracold Fermi gases},
	journal = {Reports on Progress in Physics}
}

@article{Demler_Polarons,
  title = {Fermi polarons in two dimensions},
  author = {Schmidt, Richard and Enss, Tilman and Pietil\"a, Ville and Demler, Eugene},
  journal = {Phys. Rev. A},
  volume = {85},
  issue = {2},
  pages = {021602(R)},
  numpages = {5},
  year = {2012},
  month = {Feb},
  publisher = {American Physical Society},
  doi = {10.1103/PhysRevA.85.021602},
  url = {https://link.aps.org/doi/10.1103/PhysRevA.85.021602}
}

@article{Polaron_Equations,
  title = {Attractive Fermi polarons at nonzero temperatures with a finite impurity concentration},
  author = {Hu, Hui and Mulkerin, Brendan C. and Wang, Jia and Liu, Xia-Ji},
  journal = {Phys. Rev. A},
  volume = {98},
  issue = {1},
  pages = {013626},
  numpages = {9},
  year = {2018},
  month = {Jul},
  publisher = {American Physical Society},
  doi = {10.1103/PhysRevA.98.013626},
  url = {https://link.aps.org/doi/10.1103/PhysRevA.98.013626}
}

@article{Mross_Senthil,
  title = {Controlled expansion for certain non-Fermi-liquid metals},
  author = {Mross, David F. and McGreevy, John and Liu, Hong and Senthil, T.},
  journal = {Phys. Rev. B},
  volume = {82},
  issue = {4},
  pages = {045121},
  numpages = {19},
  year = {2010},
  month = {Jul},
  publisher = {American Physical Society},
  doi = {10.1103/PhysRevB.82.045121},
  url = {https://link.aps.org/doi/10.1103/PhysRevB.82.045121}
}

@article{Hadzibabic2011,
  author = {Hadzibabic, Zoran and Dalibard, Jean},
  title = {Two-dimensional Bose fluids: An atomic physics perspective},
  journal = {La Rivista del Nuovo Cimento},
  volume = {34},
  number = {7},
  pages = {389--441},
  year = {2011},
  doi = {10.1393/ncr/i2011-10066-3},
  url = {https://springer.com}
}

@book{Pethick_Smith_Book,
  author    = {Pethick, C. J. and Smith, H.},
  title     = {Bose-Einstein Condensation in Dilute Gases},
  edition   = {2nd},
  publisher = {Cambridge University Press},
  year      = {2008},
  isbn      = {9780521846516},
  doi       = {10.1017/CBO9780511802850}
}

@article{Chubukov_anisotropic_Nematic,
  title = {Superconductivity near a nematic quantum critical point: Interplay between hot and lukewarm regions},
  author = {Klein, Avraham and Chubukov, Andrey},
  journal = {Phys. Rev. B},
  volume = {98},
  issue = {22},
  pages = {220501(R)},
  numpages = {5},
  year = {2018},
  month = {Dec},
  publisher = {American Physical Society},
  doi = {10.1103/PhysRevB.98.220501},
  url = {https://link.aps.org/doi/10.1103/PhysRevB.98.220501}
}

@article{Berg_anisotropic_Nematic,
  title = {Enhancement of Superconductivity near a Nematic Quantum Critical Point},
  author = {Lederer, S. and Schattner, Y. and Berg, E. and Kivelson, S. A.},
  journal = {Phys. Rev. Lett.},
  volume = {114},
  issue = {9},
  pages = {097001},
  numpages = {6},
  year = {2015},
  month = {Mar},
  publisher = {American Physical Society},
  doi = {10.1103/PhysRevLett.114.097001},
  url = {https://link.aps.org/doi/10.1103/PhysRevLett.114.097001}
}

@article{Chubukov_Magnetic_AFM_Critical,
doi = {10.1209/epl/i2001-00266-0},
url = {https://doi.org/10.1209/epl/i2001-00266-0},
year = {2001},
month = {may},
publisher = {},
volume = {54},
number = {4},
pages = {488},
author = {Ar. Abanov and A. V. Chubukov and A. M. Finkel'stein},
title = {Coherent vs. incoherent pairing in
2D systems  near magnetic instability},
journal = {Europhysics Letters}
}

@article{GammaI,
  author  = {Abanov, Artem and Chubukov, Andrey V.},
  title   = {Interplay between superconductivity and non-Fermi liquid at a quantum critical point in a metal. I. The $\gamma$ model and its phase diagram at $T=0$: The case $0<\gamma<1$},
  journal = {Physical Review B},
  volume  = {102},
  pages   = {024524},
  year    = {2020},
  doi     = {10.1103/PhysRevB.102.024524}
}

@article{GammaII,
  author  = {Wu, Yi-Ming and Abanov, Artem and Wang, Yuxuan and Chubukov, Andrey V.},
  title   = {Interplay between superconductivity and non-Fermi liquid at a quantum critical point in a metal. II. The $\gamma$ model at a finite $T$ for $0<\gamma<1$},
  journal = {Physical Review B},
  volume  = {102},
  pages   = {024525},
  year    = {2020},
  doi     = {10.1103/PhysRevB.102.024525}
}

@article{GammaIII,
  author  = {Wu, Yi-Ming and Abanov, Artem and Chubukov, Andrey V.},
  title   = {Interplay between superconductivity and non-Fermi liquid behavior at a quantum critical point in a metal. III. The $\gamma$ model and its phase diagram across $\gamma=1$},
  journal = {Physical Review B},
  volume  = {102},
  pages   = {094516},
  year    = {2020},
  doi     = {10.1103/PhysRevB.102.094516}
}

@article{GammaIV,
  author  = {Wu, Yi-Ming and Zhang, Shang-Shun and Abanov, Artem and Chubukov, Andrey V.},
  title   = {Interplay between superconductivity and non-Fermi liquid at a quantum critical point in a metal. IV. The $\gamma$ model and its phase diagram at $1<\gamma<2$},
  journal = {Physical Review B},
  volume  = {103},
  pages   = {024522},
  year    = {2021},
  doi     = {10.1103/PhysRevB.103.024522}
}

@article{GammaV,
  author  = {Wu, Yi-Ming and Zhang, Shang-Shun and Abanov, Artem and Chubukov, Andrey V.},
  title   = {Interplay between superconductivity and non-Fermi liquid behavior at a quantum-critical point in a metal. V. The $\gamma$ model and its phase diagram: The case $\gamma=2$},
  journal = {Physical Review B},
  volume  = {103},
  pages   = {184508},
  year    = {2021},
  doi     = {10.1103/PhysRevB.103.184508}
}

@article{GammaVI,
  author  = {Zhang, Shang-Shun and Wu, Yi-Ming and Abanov, Artem and Chubukov, Andrey V.},
  title   = {Interplay between superconductivity and non-Fermi liquid at a quantum critical point in a metal. VI. The $\gamma$ model and its phase diagram at $2<\gamma<3$},
  journal = {Physical Review B},
  volume  = {104},
  pages   = {144509},
  year    = {2021},
  doi     = {10.1103/PhysRevB.104.144509}
}

@article{MFL,
  title = {Phenomenology of the normal state of Cu-O high-temperature superconductors},
  author = {Varma, C. M. and Littlewood, P. B. and Schmitt-Rink, S. and Abrahams, E. and Ruckenstein, A. E.},
  journal = {Phys. Rev. Lett.},
  volume = {63},
  issue = {18},
  pages = {1996--1999},
  numpages = {0},
  year = {1989},
  month = {Oct},
  publisher = {American Physical Society},
  doi = {10.1103/PhysRevLett.63.1996},
  url = {https://link.aps.org/doi/10.1103/PhysRevLett.63.1996}
}

@article{Eliashberg_Limits,
  author    = {M. V. Sadovskii},
  title     = {Limits of Eliashberg theory and bounds for superconducting transition temperature},
  journal   = {Physics-Uspekhi},
  volume    = {65},
  number    = {7},
  pages     = {735--747},
  year      = {2022},
  doi       = {10.3367/UFNe.2021.05.039007},
  url       = {https://ufn.ru/en/articles/2022/7/f/}
}

@article{Chowdhury_Senthil,
  title = {Translationally Invariant Non-Fermi-Liquid Metals with Critical Fermi Surfaces: Solvable Models},
  author = {Chowdhury, Debanjan and Werman, Yochai and Berg, Erez and Senthil, T.},
  journal = {Phys. Rev. X},
  volume = {8},
  issue = {3},
  pages = {031024},
  numpages = {38},
  year = {2018},
  month = {Jul},
  publisher = {American Physical Society},
  doi = {10.1103/PhysRevX.8.031024},
  url = {https://link.aps.org/doi/10.1103/PhysRevX.8.031024}
}

@article{RPA_V_Source,
  title = {Coulomb drag of excitons in Bose-Fermi systems},
  author = {Boev, M. V. and Kovalev, V. M. and Savenko, I. G.},
  journal = {Phys. Rev. B},
  volume = {99},
  issue = {15},
  pages = {155409},
  numpages = {10},
  year = {2019},
  month = {Apr},
  publisher = {American Physical Society},
  doi = {10.1103/PhysRevB.99.155409},
  url = {https://link.aps.org/doi/10.1103/PhysRevB.99.155409}
}

@article{Extended_T_Matrix,
  title={Many Fermi polarons at nonzero temperature},
  author={Tajima, Hiroyuki and Uchino, Shun},
  journal={New Journal of Physics},
  volume={20},
  number={7},
  pages={073048},
  year={2018},
  publisher={IOP Publishing},
  doi={10.1088/1367-2630/aad1e7},
  url={https://iopscience.iop.org/article/10.1088/1367-2630/aad1e7}
}

@book{Leggett2006quantum,
  title     = {Quantum Liquids},
  author    = {Leggett, Anthony J.},
  year      = {2006},
  publisher = {Oxford University Press},
  address   = {Oxford},
  isbn      = {978-0198526438}
}

@article{Pethick_GMB,
  title = {Influence of Induced Interactions on the Superfluid Transition in Dilute Fermi Gases},
  author = {Heiselberg, H. and Pethick, C. J. and Smith, H. and Viverit, L.},
  journal = {Phys. Rev. Lett.},
  volume = {85},
  issue = {12},
  pages = {2418--2421},
  numpages = {0},
  year = {2000},
  month = {Sep},
  publisher = {American Physical Society},
  doi = {10.1103/PhysRevLett.85.2418},
  url = {https://link.aps.org/doi/10.1103/PhysRevLett.85.2418}
}

@book{altland_simons_2023,
  title     = {Condensed Matter Field Theory},
  author    = {Altland, Alexander and Simons, Ben D.},
  edition   = {3},
  publisher = {Cambridge University Press},
  year      = {2023},
  doi       = {10.1017/9781108781244},
  isbn      = {9781108494601}
}

@article{Density_Fluctuations,
  title = {Quantum field theory of dilute homogeneous Bose-Fermi mixtures at zero temperature: General formalism and beyond mean-field corrections},
  author = {Albus, A. P. and Gardiner, S. A. and Illuminati, F. and Wilkens, M.},
  journal = {Phys. Rev. A},
  volume = {65},
  issue = {5},
  pages = {053607},
  numpages = {15},
  year = {2002},
  month = {Apr},
  publisher = {American Physical Society},
  doi = {10.1103/PhysRevA.65.053607},
  url = {https://link.aps.org/doi/10.1103/PhysRevA.65.053607}
}
\end{document}